\documentclass[fleqn,usenatbib,useAMS,twocolumn]{aastex63}
\usepackage{epstopdf}
\usepackage{multimedia}
\usepackage{epsfig,graphics,subfigure,psfrag,amsmath,amssymb,amscd}
\usepackage{url, hyperref, fancyhdr}

\usepackage{acronym}
\usepackage{CJK}
\usepackage{ulem, cancel} 

\def\apjl{Astrophys. J. Lett.}
\def\apjs{Astrophys. J. Suppl.}

\def\aap{Astron. Astrophys.}

\usepackage{lineno}

\let\oldequation\equation
\let\oldendequation\endequation
\renewenvironment{equation}{\linenomathNonumbers\oldequation}{\oldendequation\endlinenomath}

\let\oldalign\align
\let\oldendalign\endalign
\renewenvironment{align}{\linenomathNonumbers\oldalign}{\oldendalign\endlinenomath}

\let\oldgather\gather
\let\oldendgather\endgather
\renewenvironment{gather}{\linenomathNonumbers\oldgather}{\oldendgather\endlinenomath}

\newcommand{\KIAA}{Kavli Institute for Astronomy and Astrophysics, Peking University, 
Beijing 100871,  People's Republic of China. 
\href{Corresponding author.}{xian.chen@pku.edu.cn}}
\newcommand{\DOA}{Department of Astronomy, School of Physics, Peking University, Beijing 100871,  People's Republic of China.}
\newcommand{\TRC}{MOE Key Laboratory of TianQin Mission, TianQin Research Center for Gravitational Physics, \\ Frontiers Science Center for TianQin, CNSA Research Center for Gravitational Waves, Sun Yat-sen University (Zhuhai Campus), Zhuhai 519082, People's Republic of China.
\href{Corresponding author.}{fanhm3@mail.sysu.edu.cn},
\href{Corresponding author.}{huyiming@mail.sysu.edu.cn}}
\newcommand{\SPA}{School of Physics and Astronomy, Sun Yat-sen University (Zhuhai Campus), Zhuhai 519082, People's Republic of China.}

\renewcommand\thetable{\Roman{table}}
\newcommand{\D}{\mathrm{d}}

\acrodef{GW}{gravitational-wave}
\acrodef{EM}{electromagnetic}
\acrodef{FLRW}{Friedmann-Lema\^itre-Robertson-Walker}
\acrodef{LCDM}[$\Lambda$CDM]{Lambda cold dark matter}
\acrodef{H0}{Hubble-Lema\^itre Constant}
\acrodef{SNIa}[SN Ia]{type Ia supernova}
\acrodefplural{SNIa}[SNe Ia]{type Ia supernovae}
\acrodef{CMB}{cosmic microwave background}
\acrodef{StBH}{stellar-mass black hole}
\acrodef{StBBH}{stellar-mass binary black hole}
\acrodef{BNS}{binary neutron star}
\acrodef{NSBHB}{neutron star-black hole binary}
\acrodefplural{NSBHB}{neutron star-black hole binaries}
\acrodef{sGRB}{short $\gamma$-ray burst}
\acrodef{HCS}{heliocentric coordinate system}
\acrodef{SNR}{signal-to-noise ratio}
\acrodef{FIM}{Fisher information matrix}
\acrodef{AGN}{active galactic nucleus}
\acrodefplural{AGN}{active galactic nuclei}

\usepackage{orcid}

\begin{document}

\begin{CJK*}{UTF8}{gbsn} 


\title{
Improving the Cosmological Constraints by Inferring the Formation Channel of Extreme-mass-ratio Inspirals
}

\author{Liang-Gui Zhu ({\CJKfamily{gbsn}朱良贵})
\orcidlink{0000-0001-7688-6504} 
}
\thanks{Boya fellow}
\affiliation{\KIAA}

\author{Hui-Min Fan ({\CJKfamily{gbsn}范会敏}) 
\orcidlink{0000-0001-5985-6178} 
}
\affiliation{Department of Physics, College of Physical Science $\&$ Technology, Hebei University, Baoding, 071002, People's Republic of China.}
\affiliation{Hebei Key Laboratory of High-precision Computation and Application of Quantum Field Theory, Baoding, 071002, People's Republic of China.}
\affiliation{Hebei Research Center of the Basic Discipline for Computational Physics, Baoding, 071002, People's Republic of China.}
\affiliation{\TRC}

\author{Xian Chen ({\CJKfamily{gbsn}陈弦})
\orcidlink{0000-0003-3950-9317} 
}
\affiliation{\KIAA}
\affiliation{\DOA}

\author{Yi-Ming Hu ({\CJKfamily{gbsn}胡一鸣})
\orcidlink{0000-0002-7869-0174} 
}
\affiliation{\TRC}
\affiliation{\SPA}

\author{Jian-dong Zhang ({\CJKfamily{gbsn}张建东}) 
\orcidlink{0000-0002-5930-6739} 
}
\affiliation{\TRC}
\affiliation{\SPA}

\date{\today}
%
\begin{abstract}

Extreme-mass-ratio inspirals (EMRIs) could be detected by space-borne
gravitational-wave (GW) detectors, such as the Laser Interferometer Space
Antenna (LISA), TianQin and Taiji.  Localizing EMRIs by GW detectors can help
us select candidate host galaxies, which can be used to infer the cosmic
expansion history.  In this paper, we demonstrate that the localization
information can also be used to infer the formation channel of EMRIs, and hence
allow us to extract more precisely the redshift probability distributions.
By conducting mock observations of the EMRIs which can be detected by TianQin and LISA, 
as well as the galaxies
which can be provided by the future Chinese Space Station Telescope, we
find that TianQin can constrain the Hubble-Lema\^itre constant $H_0$ to a
precision of $\sim3\%-8\%$ and the dark energy equation of state parameter
$w_0$ to $\sim10\%-40\%$. The TianQin+LISA network, by increasing the
localization accuracy, can improve the precisions of $H_0$ and $w_0$ to
$\sim0.4\%-7\%$ and $\sim4\%-20\%$, respectively.  Then, 
considering an illustrative case in which all EMRIs originate in AGNs,
and combining the mock EMRI observation with a mock AGN catalog, 
we show that TianQin can recognize the EMRI-AGN
correlation with $\sim 1300$ detections.
The
TianQin+LISA network can reduce this required number to $\sim 30$.
Additionally, we propose a statistical method to directly estimate the fraction
of EMRIs produced in AGNs, $f_{\rm agn}$, and show that observationally
deriving this value could significantly improve the constraints on the
cosmological parameters.  These results demonstrate the potentials of using
EMRIs as well as galaxy and AGN surveys to improve the constraints on cosmological parameters
and the formation channel of EMRIs. 

\end{abstract}


\section{Introduction}     \label{sec:introduction}

The first three observing runs of the Advanced LIGO and Virgo, and KAGRA
detectors (LVK) have detected about $90$ gravitational-wave (GW) events
triggered by the mergers of compact-object binaries \citep{2016PhRvL.116f1102A,
2017PhRvL.119p1101A, 2019PhRvX...9c1040A, 2021PhRvX..11b1053A,
2023PhRvX..13d1039A}.  In addition, evidence for the existence of GW signal in
the nanohertz band is accumulating thanks to the observations of pulsar timing
arrays \citep{2023ApJ...951L...8A, 2023A&A...678A..50E, 2023ApJ...951L...6R,
2023RAA....23g5024X}.  These observations opened a new era of observational
astronomy, in which GW can be used as a new probe of fundamental physics
\citep{2021PhRvL.126x1102A, 2021arXiv211206861T}, astrophysics
\citep{2017ApJ...851L..25F, 2020ApJ...900L..13A, 2021ApJ...912...98F,
2023PhRvX..13a1048A, 2017NatCo...8..831B, 2022MNRAS.514.2092V,
2023MNRAS.526.6031V}, and the cosmic expansion history
\citep{2017Natur.551...85A, 2019ApJ...876L...7S, 2019ApJ...871L..13F,
2019NatAs...3..940H, 2020ApJ...900L..33P, 2020ApJ...902..149V,
2021ApJ...909..218A, 2021ApJ...908..200W, 2021JCAP...08..026F,
2023ApJ...949...76A}.

The idea of combining a GW event and its electromagnetic (EM) counterpart to
establish a ``standard siren'' to probe the expansion history of the universe,
such as measuring the Hubble-Lema\^itre constant, was first proposed by
\cite{1986Natur.323..310S}.  The idea was followed by many later works which
gradually established a methodological framework even before the detection of
GW \citep{1993PhRvD..48.4738M, 2005ApJ...629...15H, 2006PhRvD..74f3006D,
2009CQGra..26i4021A, 2011ApJ...732...82P,
2011CQGra..28k4001B,2012PhRvD..85b3535T, 2012PhRvD..86d3011D,
2013arXiv1307.2638N}. The first successful measurement of the Hubble-Lema\^itre
constant using standard siren was accomplished in 2017
\citep{2017Natur.551...85A} immediately after the detection of a binary neutron
star merger in both GWs \citep{2017PhRvL.119p1101A} and EM waves
\citep{2017ApJ...848L..12A}.  For other GW events, however, a unique EM
counterpart is difficult to identify, and hence the redshift should be inferred
from the statistical distribution of the galaxies in the same sky area
localized by GW detectors \citep{1986Natur.323..310S, 2008PhRvD..77d3512M,
2011ApJ...732...82P, 2012PhRvD..86d3011D}.  Such ``dark sirens'' also provide
important constraints on the Hubble-Lema\^itre constant, e.g., in the cases of
GW170814 \citep{2019ApJ...876L...7S} and GW190814 \citep{2020ApJ...900L..33P,
2020ApJ...902..149V}.

So far, using the third Gravitational-wave Transient Catalog (GWTC-3), the LVK
collaboration have measured the Hubble-Lema\^itre constant to a precision of
$10\%$ \citep[][without using the information from the afterglow of
GW170817]{2023ApJ...949...76A}.  Better constraint can be achieved by the
next-generation ground-based GW detectors, such as the Advanced LIGO A+/Voyager
\citep{2020LRR....23....3A}, Virgo Plus \citep{2023JPhCS2429a2040A}, the
Einstein Telescope (ET) \citep{2010CQGra..27s4002P}, and Cosmic Explorer (CE)
\citep{2019BAAS...51g..35R}.  Then we can infer not only the cosmic expansion
history, but also the fractional density parameters of the universe and the
equation of state (EoS) of the dark energy \citep{2017PhRvD..95d3502D,
2017PhRvD..95d4024C, 2018PhRvD..97f4031Z, 2019JCAP...08..015B,
2020JCAP...03..051J, 2020MNRAS.498.1786Y, 2021ApJ...908..215Y,
2021ApJ...908L...4C, 2022MNRAS.512.4231B, 2022PhRvD.105b3523L,
2022arXiv221200531S}.  These measurements will help solve the most difficult
questions in today's cosmology, such as the inconsistency between the measured
values of the Hubble-Lema\^itre constant in the early and late universe
\citep{2019PhRvL.122f1105F, 2020ApJ...905L..28B, 2023PhRvD.107l3519C,
2023MNRAS.524.3537G}, which is known as the ``Hubble tension''
\citep{2017NatAs...1E.121F, 2020A&A...641A...6P, 2021ApJ...908L...6R, 
2021CQGra..38o3001D, 2022NewAR..9501659P}. 

In about a decade, we are likely to have space-borne GW detectors, such as the
Laser Interferometer Space Antenna (LISA) \citep{2017arXiv170200786A,
2024arXiv240207571C}, TianQin \citep{2016CQGra..33c5010L, 2021PTEP.2021eA107M},
or Taiji \citep{Hu:2017mde}.  These detectors are sensitive to the GWs in the
band of $0.1 ~\!{\rm mHz} -  1 ~\!{\rm Hz}$.  In this band, the potential GW
sources which can be used as standard sirens include stellar-mass
($\sim10M_\odot$) binary black holes (BHs), massive BH (MBH) binaries
($10^5-10^8M_\odot$), and extreme-mass-ratio inspirals (EMRIs) \citep[see][for
a review]{2023LRR....26....5A}.  The usage of the first two sources in
cosmology have been extensively studied \citep{2005ApJ...629...15H,
2011ApJ...732...82P, 2016JCAP...04..002T, 2016JCAP...10..006C,
2018MNRAS.475.3485D, Wang:2020dkc, 2022SCPMA..6510411W, 2022PhRvR...4a3247Z,
2022SCPMA..6559811Z, 2022PhRvD.105d3509M, 2023SCPMA..6720412J}.  These studies
show that the stellar-mass binary BHs detectable by space-borne GW detectors are
predominately in the local universe \citep{2016PhRvL.116w1102S,
2016MNRAS.462.2177K, 2020PhRvD.101j3027L}, while MBH binaries (MBHBs) are among
the loudest GW sources in the milli-Hertz band and hence can be detected to
high redshift \citep[$z\ga10$,][]{2005ApJ...629...15H, 2007PhRvD..76j4018C,
2016PhRvD..93b4003K, 2017arXiv170200786A, 2019PhRvD.100d3003W}.

The third type of low-frequency GW sources which can be used as standard sirens
are EMRIs. EMRI normally resides at the center of a galaxy and consists of an
MBH and an inspiralling stellar-mass BH \citep{2018LRR....21....4A}.  The
potential of using LISA EMRIs to study cosmology was first pointed out in
\cite{2008PhRvD..77d3512M}, and later studied in more detail in
\cite{2021MNRAS.508.4512L, 2023arXiv231012813L}.  These earlier studies
considered only the EMRIs with $z < 1$ where the galaxy catalogs are relatively
more complete.  However, population studies have shown that space-borne GW
detectors can catch EMRIs up to $z \approx 3$ or higher
\citep{2017PhRvD..95j3012B, 2020PhRvD.102f3016F, 2020PhRvD.102j3023B}.  At such
high redshifts, although the current galaxy catalog is incomplete, future space
telescopes will be able to conduct a more complete survey of the galaxies
\citep{2019ApJ...883..203G, 2019ApJ...873..111I, 2022A&A...662A.112E}.  In
addition, besides LISA, other space-borne GW detectors, such as TianQin and
Taiji, can also detect EMRIs. A joint observation by multiple detectors in
principle can improve the sky localization of a GW source by orders of
magnitude \citep{2020NatAs...4..108R, 2021NatAs...5..881G}.  These
observational prospects in general should improve the constraint on
cosmological parameters by EMRIs.

The prospect of using EMRIs to constrain the cosmological parameters relies on
not only the completeness of a galaxy catalog, but also the accuracy of the
redshift which can be inferred from the galaxy catalog. When extracting the
redshift probability distributions of EMRIs from galaxy catalogs, previous
works \citep{2008PhRvD..77d3512M, 2021MNRAS.508.4512L, 2023arXiv231012813L}
have assumed that all galaxies have an equal probability to host an EMRI.
However, there might exist a substantial bias towards specific types of
galaxies in the formation of EMRIs, as alternative formation channels could
potentially yield higher EMRI rates in comparison to the standard channel
\citep{2005ApJ...629..362H, 2017PhRvD..95j3012B, 2018LRR....21....4A}.  These
alternative channels include binary separations \citep{2005ApJ...631L.117M},
MBHB systems \citep{2022MNRAS.516.1959M, 2022ApJ...927L..18N,
2023ApJ...955L..27N} and AGNs \citep{2003astro.ph..7084L, 2007MNRAS.374..515L,
2011PhRvL.107q1103Y,2021PhRvD.103j3018P, 2021PhRvD.104f3007P,
2023MNRAS.521.4522D}.  In particular,  AGNs are potentially favourable
environments for forming EMRIs because the accretion disks of AGNs can capture
more stellar-mass BHs and BH binaries and drive them to migrate towards the
central MBHs than normal galactic nuclei can do \citep{2020ApJ...898...25T,
2023PhRvX..13a1048A}.  The major properties predicted by the different EMRI
formation channels, such as the EMRI rates, eccentricities at plunges, and
masses of the MBHs and stellar-mass BHs, are subject to considerable
uncertainties, posing challenges in determining the origins of EMRIs. 
Therefore, what type of galaxies in a catalog can be considered as the hosts of
EMRIs remains a question. 

Recently, a statistical framework has been proposed  which allows one to test
the spatial correlation between GW sources and AGNs even in the absence of
identifying the host galaxies of the GW sources \citep{2017NatCo...8..831B}.
The framework utilizes a Poisson distribution to describe the likelihood of the
number of candidate host AGNs for the GW source, and constrains the spatial
correlation between the GW sources and AGNs by a likelihood-ratio-based method.
Several works have demonstrated the effectiveness of this framework in
identifying the host galaxies of the LVK binary BHs \citep{2022MNRAS.514.2092V,
2023MNRAS.526.6031V} or the MBHBs of future space-borne GW detectors
\citep{2024ApJ...960...43Z}.  Such a method in principle can be used to infer
the fraction of EMRIs which originate in AGNs.  Knowing this fraction will help
us better weigh the large number of candidate host galaxies in the sky area
localized for an EMRI, so that we can more precisely extract the redshift
probability distributions and measure cosmological parameters.

Therefore, in this work we will apply the aforementioned statistical framework
to EMRIs, and study its effectiveness in constraining the cosmological
parameters and testing the spatial correlation between EMRIs and AGNs.  To
facilitate the study, we consider TianQin, as well as a network composed of
TianQin and LISA, as the future observational instruments.  This paper is
organized as follows.  In Section~\ref{sec:method} we introduce the statistical
analysis framework that we use to infer the cosmic expansion history and test
the EMRI-AGN correlation.  In Section \ref{sec:data} we conduct mock
observations.  The results are shown in Section~\ref{sec:result_cosmo} for the
cosmological prospects and Section~\ref{sec:result_astrop} for the prospects of
inferring the EMRI-AGN correlation.  In Section~\ref{sec:discussion}, we
discuss some of the factors that can affect the results.  Finally, in
Section~\ref{sec:conclusion} we summarize the main findings of this work and
the implications of our results.

\section{Methodology}     \label{sec:method}

\subsection{Cosmological model}    \label{sec:cosmo_model}

We assume that the universe is flat \citep{2020A&A...641A...6P, 2021MNRAS.506L...1D}
and can be described by the Friedmann-Lema\^itre-Robertson-Walker metric. 
From the Friedmann equations, the expansion rate of the universe can be described 
by 
\begin{align}
\!\!\!\!\!\!\!\!\!\!\!\!\!  H(z) = H_0 \sqrt{\Omega_M(1 \! + \! z)^3 + \Omega_\Lambda \exp \!\left[3 \!\! \int_0^z \! \frac{1 \! + \! w(z')}{1 \! + \! z'} \D z' \right] },  \label{eq:H_z}
\end{align}
where $H(z)$ is called as
the Hubble-Lema\^itre parameter and $H_0 \equiv H(z=0)$ is its current value, 
$\Omega_M$ and $\Omega_\Lambda$ are respectively the ratios of the total matter density 
and dark energy density to the critical density of the universe 
(they satisfy $\Omega_M + \Omega_\Lambda = 1$ in a flat-universe), and $w(z)$ describes 
the dark energy EoS as a function of redshift $z$. 

In this work, we consider two dark energy models, 
namely the standard model---the cosmological constant $\Lambda$, 
and the Chevallier-Polarski-Linder (CPL) model \citep{2001IJMPD..10..213C, 2003PhRvL..90i1301L}. 
(i) For the cosmological constant model, the EoS of the dark energy is constantly equal to minus one, 
i.e., $w(z) \equiv -1$. This model, together with 
the cold dark matter model and the spatially-flat assumption, 
constitute the so-called standard cosmological model---flat-$\Lambda$CDM. 
(ii) For the CPL model, the EoS of the dark energy is modeled as a function
of $w (z) = w_0 + w_a {z}/{(1+z)}$ 
\citep{2001IJMPD..10..213C}, where $w_0$ and $w_a$ are two extra parameters. 
The analytic expression of Equation (\ref{eq:H_z}) is
$H(z) = H_0 \sqrt{\Omega_M (1 \! + \! z)^3 + \Omega_{\Lambda} (1 \! + \! z)^{3(1+w_0+w_a)} \exp \! \big( \!\! - \! \frac{3w_a z}{1+z} \! \big)}$. 
In particular, when $w_0 = -1$ and $w_a = 0$, the CPL model 
recovers the cosmological constant model. 

Regardless of dark energy models, 
we have the ``luminosity distance$-$redshift'' ($D_L - z$) relation under a flat-universe according to the expansion effect of the universe, i.e., 
\begin{equation}  \label{eq:DL-z} 
D_L = c(1+z) \! 
\int_{0}^{z} \! \frac{1}{H(z')} \D z' , 
\end{equation} 
where $c$ is the speed of light.  In our simulations, we fix the cosmological
parameters to the values: $H_0 \equiv h \times 100 ~\! {\rm km} ~\! {\rm
s}^{-1} ~\!\! {\rm Mpc}^{-1} = 67.8 ~\! {\rm km} ~\! {\rm s}^{-1} ~\!\! {\rm
Mpc}^{-1}$ (here $h$ is the dimensionless Hubble-Lema\^itre constant),
$\Omega_M = 0.307$, $w_0 = -1$, and $w_a = 0$.

\subsection{Bayesian framework for GW cosmology}    \label{sec:Bayesian_framework}

To constrain the cosmological parameters by combining the GW and EM data, we
adopt the Bayesian framework presented in \cite{2018Natur.562..545C}.  However
the estimation of the completeness of the galaxy catalog is modified.
Here the estimation is based on the correlation between the 
MBHs in EMRIs and their host galaxies. More specifically, we
use the approximate linear relationship between the masses of MBHs and the
(bulge) luminosities of the galaxies \citep[which is known as the $M_{\rm
MBH}-L$ relation;][] {2000ApJ...539L...9F, 2013ApJ...763...76S,
2013ARA&A..51..511K}. 

Given a GW data set composed of $N$ independent GW events 
$\mathcal{D}^{\rm gw} \equiv \{d^{\rm gw}_1,  d^{\rm gw}_2, \dots, d^{\rm gw}_i, \dots, d^{\rm gw}_N \}$ 
as well as the corresponding data set of EM counterparts 
$\mathcal{D}^{\rm em} \equiv \{d^{\rm em}_1, d^{\rm em}_2, \dots, d^{\rm em}_i, \dots, d^{\rm em}_N \}$, 
the \emph{posterior} probability distribution of the cosmological parameters 
(denote as ${\bf \Omega} \equiv \{H_0, \Omega_M, w_0, w_a\}$)
that we are interested in can be expressed as 
\begin{align}  \label{eq:posterior} 
\!\!\!\!\!\!\!
p({\bf \Omega} |\mathcal{D}^{\rm gw}, \mathcal{D}^{\rm em}, I) \propto p_0({\bf \Omega}|I) p(\mathcal{D}^{\rm gw},
\mathcal{D}^{\rm em}|{\bf \Omega}, I)  & \nonumber \\ 
\propto p_0({\bf \Omega}|I) \prod_i p(d^{\rm gw}_i, d^{\rm em}_i|{\bf \Omega}, I) &, 
\end{align} 
where $I$ indicates all the relevant additional information, 
$p_0({\bf \Omega}|I)$ is the \emph{prior} probability distribution for ${\bf \Omega}$, 
and $p(d^{\rm gw}_i, d^{\rm em}_i|{\bf \Omega}, I)$ represents the \emph{likelihood} 
of observing the $i$th GW event as well as the corresponding EM signal. 
Because the data contain noise such that different sources with different 
right ascensions $\alpha$ and declinations $\delta$, luminosity distances $D_L$, 
redshifts $z$, masses $M$, and luminosities $L$ may generate the same observational data. 
Therefore, the likelihood needs to marginalize all possible $(\alpha, \delta, D_L, z, M, L)$, i.e.,
\begin{align} 
\label{eq:likeli_decompos}
\!\!\!\!\!\!\!\!\!\!\!\!\!
 & p(d^{\rm gw}_i, d^{\rm em}_i| {\bf \Omega}, I)  \nonumber \\
\!\!\!\!\!\!\!\!\!\!\!\!\! =&  \frac{\int\! p(d^{\rm gw}_i \!,\! d^{\rm em}_i \!,\! \alpha,\! \delta,\! D_L,\! z,\! M,\! L | {\bf \Omega},\! I) \D\alpha  \D \delta  \D D_L \D z  \D M  \D L}{\beta({\bf \Omega} | I)},
\end{align}
where $\beta({\bf \Omega} | I)$ is a normalization factor accounting for 
the selection effects in GW and EM observations 
\citep{2019MNRAS.486.1086M, 2018Natur.562..545C, 2019ApJ...876L...7S}. 

To calculate the numerator in Equation (\ref{eq:likeli_decompos}), 
we separate $p(d^{\rm gw}_i \!, d^{\rm em}_i \!, \alpha, \delta, D_L, z, M, L | {\bf \Omega}, I)$ into
\begin{align}  
\label{eq:likeli_separate}
\!\!\!\!\!\!\!\!\!\!\!\!\!
 & p(d^{\rm gw}_i, d^{\rm em}_i, \alpha, \delta, D_L, z, M, L | {\bf \Omega}, I)   \nonumber \\
\!\!\!\!\!\!\!\!\!\!\!\!\!
 = &~ p(d^{\rm gw}_i, d^{\rm em}_i |\alpha, \delta, D_L, z, M, L, {\bf \Omega}, I)   \nonumber \\
\!\!\!\!\!\!\!\!\!\!\!\!\!
 & \!\!\times p_0(\alpha, \delta, D_L, z, M, L |{\bf \Omega}, I)   \nonumber \\
\!\!\!\!\!\!\!\!\!\!\!\!\!
 = &~ p(d^{\rm gw}_i |\alpha, \delta, D_L, M , {\bf \Omega}, I) p(d^{\rm em}_i |\alpha, \delta, z, L, {\bf \Omega}, I)    \nonumber \\
\!\!\!\!\!\!\!\!\!\!\!\!\!
 & \!\!\times p_0(D_L|z, {\bf \Omega}, I)  p_0( M |z, L, {\bf \Omega}, I) 
  p_0(\alpha, \delta, z, L |{\bf \Omega}, I).
\end{align}
The term $p(d^{\rm gw}_i |\alpha, \delta, D_L, M, {\bf \Omega}, I)$ is 
the likelihood of the GW data \citep{1992PhRvD..46.5236F, 2009LRR....12....2S}
after marginalized over all other parameters that are unrelated to the cosmic expansion history, 
and $p(d^{\rm em}_i |\alpha, \delta, z, L, {\bf \Omega}, I)$ is the likelihood 
of the corresponding EM data. For dark sirens, 
because there is no information from the observations of the EM counterparts, 
we can define $d^{\rm em}_i$ to null, as well as define the likelihood 
$p(d^{\rm em}_i |\alpha, \delta, z, L, {\bf \Omega}, I)$ to constant \citep{2018Natur.562..545C}. 
The term $p_0(D_L|z, {\bf \Omega}, I)$ is a Dirac delta function, i.e., 
$p_0(D_L|z, {\bf \Omega}, I) = \delta_{\rm D} \big(D_L - \hat D_L(z, {\bf \Omega}) \big)$ 
(here $\hat D_L(z, {\bf \Omega})$ represents Equation (\ref{eq:DL-z})), 
because $D_L$ is a function of $z$ given a cosmology. 

The term $p_0( M |z, L, {\bf \Omega}, I)$ can be approximated as 
a uniform distribution of $M$, i.e., 
%
\begin{align}  
\label{eq:likeli_ML}
\!\!\!\!\!\!\!\!\!\!
& p_0( M |z, L, {\bf \Omega}, I)  \propto   \nonumber \\
\!\!\!\!\!\!\!\!\!\!
& \mathcal{U}\big[(1 \!+\! z)(\hat M(L) \!-\! 3\sigma_{\! M}), 
(1 \!+\! z)(\hat M(L) \!+\! 2\sigma_{\! M}) \big] (M), 
\end{align}  
where the two values in square bracket $\mathcal{U}[... ~\!, ...]$ represent the lower 
and upper boundaries of the uniform distribution, $\hat M(L)$ represents the mass of the MBH 
as a function of the luminosity of the host galaxy, 
and $\sigma_{\! M}$ is the standard deviation. 
The expression of $p_0( M |z, L, {\bf \Omega}, I)$ holds just for 
EMRI (and MBHB) GW sources due to the $M_{\rm MBH}-L$ relation. 
The reason for adopting a uniform rather than a Gaussian distribution 
is that the data for which the $M_{\rm MBH}-L$ relation is derived 
have many outliers \citep{2000ApJ...539L...9F, 
2013ApJ...763...76S, 2013ARA&A..51..511K}. 
The function $\hat M(L)$ can be written as 
\begin{align} \label{eq:M_L}
\!\!\!\!\!\!\!\!\!\!\!\!\!
\lg \!\left(\! \frac{\hat M(L)}{M_{\odot}} \!\right)\!  \approx   a_{\rm fit} + 
b_{\rm fit} \!\left[\! M_{\odot}^* - 2.5 \lg \!\left(\! \frac{L}{L_{\odot}} \!\right) + 23.4 \right]\! , 
\end{align}
\citep{2013ApJ...763...76S}, where $a_{\rm fit}$ and $b_{\rm fit}$ are the fitting zeropoint and slope, 
and $M_{\odot} \!=\! 1.989 \times 10^{30} ~\! {\rm kg}$, $M^*_{\odot} \!=\! +4.8 \!~{\rm mag}$ and $L_{\odot} \!=\! 3.8 \times 10^{33} \!~{\rm erg ~s}^{-1}$ 
are the mass, absolute magnitude, and luminosity of the Sun, respectively. 
The best-fitting parameters for Equation (\ref{eq:M_L}) in the $K$-band are 
$a_{\rm fit} = 8.04$ and $b_{\rm fit} = -0.48$ with 
a root-mean-square scatter of $\sigma_M = 0.4$ in $\lg (M/M_{\odot})$ \citep{2013ApJ...763...76S}.
Notice that the uniform distribution adopted by us is not symmetric 
around $\hat M(L)$ because there is a fraction of galaxies with lighter MBHs 
compared to the predicted $\hat M(L)$ \citep{2011ApJ...737L..45J, 2013ApJ...763...76S, 2013ARA&A..51..511K}. 

The last term $p_0(\alpha, \delta, z, L |{\bf \Omega}, I)$ in Equation (\ref{eq:likeli_separate}) 
is a prior distribution for the spatial positions and luminosities of 
galaxies. 
To get a proper prior, we adopt the standard assumption that 
the spatial probability distribution of GW sources is proportional to
the number density of galaxies \citep{1986Natur.323..310S, 2008PhRvD..77d3512M, 
2011ApJ...732...82P, 2012PhRvD..86d3011D, 2018Natur.562..545C, 2019ApJ...871L..13F, 
2019ApJ...876L...7S, 2020PhRvD.101l2001G, 2021JCAP...08..026F, 2023ApJ...949...76A}. 
Further taking into account the 
selection effect that many galaxies are too dim to be recorded by surveys, 
we can express the prior as
\begin{align} \label{eq:prior_totgalaxy}
\!\!\!\!\!\!\!
 p_0(\alpha, \delta, z,  L |& {\bf \Omega}, I) = ~ f_{\rm compl} ~\! p_{\rm obs}(\alpha, \delta, z, L |{\bf \Omega}, I)  \nonumber \\
&+  (1 - f_{\rm compl}) ~\! p_{\rm miss}(\alpha, \delta, z, L |{\bf \Omega}, I),
\end{align}
where $p_{\rm obs}(\alpha, \delta, z, L |{\bf \Omega}, I)$ represents a
distribution function of observed galaxies from EM surveys, 
$f_{\rm compl}$ is the completeness fraction of the observed galaxy catalog, 
and $p_{\rm miss}(\alpha, \delta, z, L |{\bf\Omega}, I)$ represents the 
probability distribution function (PDF) of the unobservable galaxies. 

To calculate $p_0(\alpha, \delta, z,  L |{\bf \Omega}, I)$, we need specific expressions 
for the two terms on the right-hand side of Equation (\ref{eq:prior_totgalaxy}).
(i) The distribution function of observed galaxies $p_{\rm obs}(\alpha, \delta, z, L |{\bf \Omega}, I)$ 
can generally be written as
\begin{align} \label{eq:prior_obsgalaxy}
\!\!\!\!\!\!\!\!\!\!\!\!\!\!
 p_{\rm obs}(\alpha, \delta, z, L | {\bf \Omega}, & I) \propto \sum_{j=1}^{N_{\rm gal}} 
 \Big[ \delta_{\rm D} (\alpha - \alpha_j) ~\! \delta_{\rm D} (\delta - \delta_j)  \nonumber \\
 & ~~ \times \! \mathcal{N}[\bar z_j, \sigma_{z;j}](z) ~\! 
  \mathcal{N}[\bar L_j, \sigma_{L;j}](L) \Big], 
\end{align}
where $N_{\rm gal}$ is the total number of the observed galaxies and two 
$\mathcal{N}[\bar x, \sigma_x](x)$ represent Gaussian distributions of $x$ 
with mean values $\bar x$ and standard deviations $\sigma_x$. 
Here, because the errors of sky positions of galaxies in EM surveys are much smaller than 
the sky localization errors of the GW sources, 
the distribution of $(\alpha, \delta)$ of galaxies 
can be approximately expressed as two Dirac delta functions. 
(ii) For the unobservable galaxies, their PDF can be expressed as
\begin{align}
\!\!\!\!\!\!\!\!\!\!\!
p_{\rm miss}(\alpha, \delta, z, L | {\bf \Omega}, I) \propto  
{\rm erfc} (L) ~\!  \Phi(L, z) 
 \frac{\D V_{\rm c}}{\D z \D \hat\Omega} \D z \D \hat\Omega ,  \label{eq:prior_missgalaxy}
\end{align}
where   
\begin{align}
\!\!\!\!\!\!\!\!\!\!
 {\rm erfc} (L) =  \frac{1}{\sqrt{2\pi \bar\sigma_{L}^2}} \!\int_{L}^{\infty} \!\! \exp\!\Bigg[\!\! -\frac{\big(L' - \hat L_{\rm thr}(z, {\bf \Omega}) \big)^2 }{2 ~\! \bar\sigma_{L}^2}\Bigg] \! \D L'    \nonumber 
\end{align}
is the complementary error function with mean value $\hat L_{\rm thr}(z, {\bf \Omega})$ 
and standard distribution $\bar\sigma_{L}$,  
$\Phi(L,z)$ represents the luminosity function of the galaxies at $z$ 
\citep{1976ApJ...203..297S, 2005ApJ...631..126D, 2007ApJ...656...42M}, 
$V_{\rm c}$ is the comoving volume, and 
$\D \hat\Omega = \cos \delta \!~ \D \alpha \D \delta$ represents the differential solid angle. 
The mean value $\hat L_{\rm thr}(z, {\bf \Omega})$ is a luminosity threshold 
for an observable galaxy with $z$ in a cosmology ${\bf \Omega}$, and it depends on 
the limiting magnitude $m_{\rm limit}^*$ of the galaxy survey. We can calculate it with 
\begin{align}  
\label{eq:appMag_mi}
\!\!\!\!\!\!\!\!\!\!  
 \lg \!\left(\! \frac{\hat L_{\rm thr}(z, {\bf \Omega})}{L_{\odot} } \!\right)\!  = & ~\!
  \frac{1}{2.5} \Bigg[ M^*_{\odot} - m_{\rm limit}^*   \nonumber \\
\!\!\!\!\!\!\!\!\!\!
&  + 5\lg \!\left(\! \frac{\hat D_L(z, {\bf \Omega})}{1 ~\!{\rm Mpc}}  \!\right)\! + 25  \Bigg].
\end{align}

The complete expression of $p_0(\alpha, \delta, z,  L |{\bf \Omega}, I)$ 
additionally requires a reasonable estimation for the completeness of 
the observed galaxy catalog.
In this work, we consider the role of the $M_{\rm MBH}-L$ relation 
for estimating the completeness of galaxy catalogs.
We assume that the completeness  
depends not only on the galaxy catalog itself, 
but also on the property of the EMRI source (such as the mass of the MBH). 
The completeness fraction can be estimated by 
\begin{align} \label{eq:f_compl}
\!\!\!\!\!\!\!\!\!\!\!\!\!\!
 f_{\rm compl} & \approx  1 \! - \! \bigg[ \! \int \!\! p(d^{\rm gw}_i |\alpha, \delta, D_L, M, {\bf \Omega}, I) 
 p_0(D_L|z, {\bf \Omega}, I)  \nonumber \\
\!\!\!\!\!\!\!\!\!\!\!\!\!\!
& \times p_0( M |z, L, {\bf \Omega}, I)     \nonumber \\
\!\!\!\!\!\!\!\!\!\!\!\!\!\!
& \times p_{\rm miss}(\alpha, \delta, z, L | {\bf \Omega}, I) \D\alpha \D\delta \D D_L \D M \D z \D L \bigg].
\end{align}
The $M_{\rm MBH}-L$ relation is the statistical basis that enables the term 
$p_0( M |z, L, {\bf \Omega}, I)$ to connect the EMRI GW source to the galaxy catalog. 
From Equation (\ref{eq:f_compl}), we can find that 
for an EMRI source with a more massive MBH, 
the completeness of the same galaxy catalog is higher. 

Finally, to calculate the denominator $\beta({\bf \Omega} | I)$ in Equation (\ref{eq:likeli_decompos}), 
we adopt the same estimated approach as presented in 
\cite{2018Natur.562..545C} 
\citep[also see][]{2020PhRvD.101l2001G, 2021JCAP...08..026F, 
2023ApJ...949...76A, 2023ApJ...948...26Z}.  
For dark sirens, the denominator is determined by 
the GW data $d^{\rm gw}$ and the prior $p_0(\alpha, \delta, z,  L |{\bf \Omega}, I)$,  
and we have 
\begin{align} \label{eq:belta_term}
\!\!\!\!\!\!\!\!\!\!\!\!
 \beta({\bf \Omega} | I) &= \!\!\! \int\limits_{{\rm detectable}}  \!\!\!\!\!\!  p(d^{\rm gw}|  \hat D_L(z, {\bf \Omega}), M, I)  p_0( M |z, L, {\bf \Omega}, I)    \nonumber \\
\!\!\!\!\!\!\!\!\!\!\!\!
& \!\!\! \times    p_0(\alpha, \delta, z, L | {\bf \Omega}, I) ~\! \D d^{\rm gw} \D \alpha \D \delta \D M \D z.  
\end{align}
Here, the equation represents an integration over the possible GW data $d^{\rm
gw}$ that the signal-to-noise ratio (SNR) is greater than the detection
threshold.

\subsection{Statistical framework for constraining the formation channel of EMRIs}    \label{sec:Poisson_framework}

An effective constraint on the formation channel of EMRIs can be obtained by 
testing the spatial correlation between EMRIs and AGNs. 
We set the null hypothesis of our test to be that EMRIs can reside in either normal galaxies or AGNs. 
It means that the probability that a detected EMRI originates in an AGN 
is proportional to the fraction of AGNs among all galaxies. 
Our alternative hypothesis is that EMRIs predominantly reside in AGNs. 

To statistically test our null hypothesis, 
we use the same statistical framework presented in \cite{2024ApJ...960...43Z}, 
which was developed upon the work of \cite[][also see earlier \cite{2008APh....29..299B}]{2017NatCo...8..831B} 
and modified to test the spatial correlation between 
MBHB mergers and AGNs. 
Given that the spatial number density of AGNs is $\rho_{\rm agn}$, 
and the spatial localization error volume of the $i$th detected EMRI is $\Delta V_i$,
in the null hypothesis the number $N_{{\rm agn},i}$ of AGNs within the error volume $\Delta V_i$ 
will follow a Poisson distribution with an expectation of $\rho_{\rm agn} \Delta V_i$, i.e.,
\begin{align}  \label{eq:Poisson-Bi} 
B_i(N_{{\rm agn},i}) = {\rm Poisson} \Big( N_{{\rm agn},i}, \rho_{\rm agn} \Delta V_i \Big).
\end{align} 
In the alternative hypothesis, there is one guaranteed host AGN within $\Delta V_i$. 
Therefore, the number of the rest non-host AGNs, $N_{{\rm agn},i} -1$, will follow 
the Poisson distribution with the expectation $\rho_{\rm agn} \Delta V_i$, i.e., 
\begin{align}  \label{eq:Poisson-Si} 
S_i(N_{{\rm agn},i}) = {\rm Poisson} \Big( N_{{\rm agn},i}-1, \rho_{\rm agn} \Delta V_i \Big).
\end{align} 

Taking into account our lack of knowledge about the formation channel of EMRIs, 
we assume that a fraction $f_{\rm agn}$ of detected EMRIs are from AGNs, 
while the rest detected EMRIs originate in normal galaxies. 
Effectively, it means that the probability that 
an individual EMRI originates in an AGN is equal to $f_{\rm agn}$. 
Furthermore, taking into account the incompleteness of the AGN catalog,
we can write the individual likelihood of the $i$th detected EMRI as 
\begin{align}  \label{eq:likeli_fagn_i} 
\!\!\!\!\!\!\!\!
\mathcal{L}_i(f_{\rm agn}) =  f_{\rm agn} f_{{\rm compl},i}^{\rm agn}  S_i 
                  + \Big(\! 1- f_{\rm agn} f_{{\rm compl},i}^{\rm agn} \!\Big) B_i 
\end{align} 
\citep{2024ApJ...960...43Z}, where $f_{{\rm compl},i}^{\rm agn}$ represents the completeness 
fraction of the AGN catalog when the $i$th detected EMRI is evaluated. 
For the estimation of $f_{{\rm compl},i}^{\rm agn}$, we follow exactly 
the approach of \cite{2024ApJ...960...43Z}, by considering the correlation 
between the luminosities of AGNs and the masses of their central MBHs. 
For a GW catalog composed of $N$ EMRIs, 
the total likelihood can be expressed as 
\begin{align}  \label{eq:likeli_fagn_tot} 
\mathcal{L}(f_{\rm agn}) = \prod_{i} \mathcal{L}_i(f_{\rm agn}) .
\end{align} 
Several factors render it exceedingly challenging to accurately infer $f_{\rm agn}$ 
directly from Equation (\ref{eq:likeli_fagn_tot}), 
including (i) the degeneracy between $f_{\rm agn}$ and $f_{{\rm compl},i}^{\rm agn}$, 
(ii) the sensitivity of the likelihood to the accurate determination of $\rho_{\rm agn}$, and
(iii) the data $\{ N_{{\rm agn}, i} \}$ of the likelihood is not directly related to $f_{\rm agn}$. 
Regarding (i) and (ii), 
accurate estimations of both $f_{{\rm compl},i}^{\rm agn}$ and $\rho_{\rm agn}$ 
are impeded by selection effects. 
For (iii), Because the number $N_{{\rm agn}}$ of candidate host AGNs is proportional to the spatial localization error volume of the EMRI, not by correlation with $f_{\rm agn}$. 
An EMRI originated in an AGN 
merely implies an anticipated increase of one in the count of candidate host AGNs for this EMRI, 
compared to a random scenario. 

To statistically constrain the fraction $f_{\rm agn}$ in Equation (\ref{eq:likeli_fagn_tot}), we follow \cite{2008APh....29..299B} and \cite{2017NatCo...8..831B} to introduce
a likelihood ratio
\begin{align}    \label{eq:likeli-ratio} 
\lambda = 2 \ln \!\left[ \frac{\mathcal{L}(f_{\rm agn})}{\mathcal{L}(0)} \right].
\end{align} 
In the null hypothesis, in which the value of $f_{\rm agn}$ equals to the
fraction of AGNs among all galaxies in the universe, the likelihood ratio
$\lambda$ is governed by the random fluctuation of the spatial number
density of AGNs, and will follow a ``background distribution'' $P_{\rm
bg}(\lambda)$. 
Conversely, in the alternative hypothesis that $f_{\rm agn}$ significantly exceeds 
the fraction of AGNs among all galaxies, $\lambda$ becomes governed by 
the spatial correlation between EMRIs and AGNs. 
This is because, under this condition, the likelihoods $\mathcal{L}(f_{\rm agn})$ and $\mathcal{L}(0)$ 
in Equation (\ref{eq:likeli-ratio}) exhibit significant discrepancies. 
For illustrative purposes, the distributions of the likelihood ratio
$\lambda$ in the null and alternative hypotheses, as well as their variation with 
$f_{\rm agn}$,
are presented in Appendix \ref{appendix:lambda_bg}.
A very important property is that the $\lambda$
derived from real observation of EMRIs 
will deviate more from $P_{\rm bg}(\lambda)$ when the fraction of 
the EMRIs originate in AGNs is larger.  In the
analysis, $P_{\rm bg}(\lambda)$ can be determined by Monte-Carlo simulations,
and the deviation of the median $\lambda$ relative to $P_{\rm bg}(\lambda)$ can be
quantified by an $p$-value. 
If this $p$-value becomes less than $0.00135$, we
can claim that we can reject the null hypothesis with a significance of
$3\sigma$, i.e., confirm the spatial correlation between the EMRIs and AGNs.

Finally, we estimate the uncertainty in the observational constraint 
on $f_{\rm agn}$.  We denote the likelihood ratio derived from
observational data (the numbers of candidate host AGNs of the detected EMRIs)
as $\lambda_{\rm s}(f_{\rm agn})$, where the subscript ``s'' stands for
``signal''. Notice that its value varies with the underlying AGN fraction $f_{\rm agn}$.
To check whether $\lambda_{\rm s}(f_{\rm agn})$ is consistent with or can be rejected
by the background distribution of  $\lambda$, we 
introduce an integrated probability
\begin{align}    \label{eq:CDF-fagn} 
CDF(\lambda_{\rm s}, f_{\rm agn}) =  \frac{1}{\mathcal{N}}  \int_{\lambda_{\rm s}(f_{\rm agn})}^{\infty} \!\!
P(\lambda | f_{\rm agn}) \D \lambda ,
\end{align} 
where $P(\lambda | f_{\rm agn})$ is the conditional PDF of the likelihood ratio
$\lambda$ given $f_{\rm agn}$, and $\mathcal{N}$ is a normalization factor.
In our simulations, each $P(\lambda | f_{\rm agn})$ is derived from about $10^4$ 
independent Monte-Carlo simulations. 

\begin{figure}[htbp]
\centering
\includegraphics[width=0.460\textwidth]{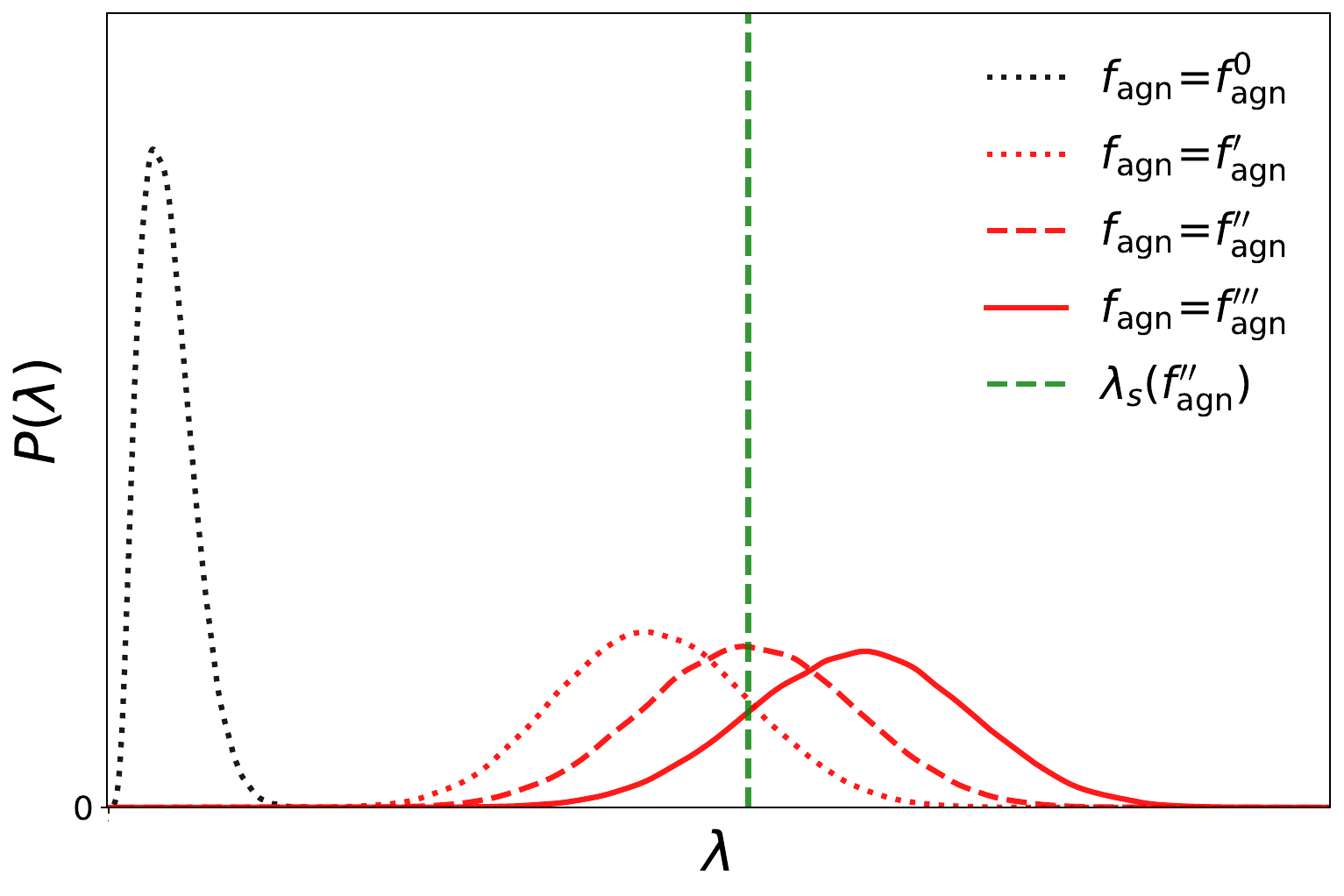}
\caption{
An example comparing the signal likelihood ratio $\lambda_{\rm s}(f_{\rm agn})$
(vertical green dashed line) with the conditional PDFs of $\lambda$ for four different values of
$f_{\rm agn}$. In this example, $f^0_{\rm agn}=0.1$, $f'_{\rm agn}=0.55$, $f''_{\rm agn}=0.7$,
and $f'''_{\rm agn}=0.85$.
}
\label{fig:fagn_estimate}
\end{figure}

The meanings of $\lambda_{\rm s}(f_{\rm agn})$ and $P(\lambda | f_{\rm agn})$
are illustrated in Figure~\ref{fig:fagn_estimate}. It shows that the conditional
PDF $P(\lambda | 0.1)$ (black dotted curve) is clearly inconsistent with the observed value of
$\lambda_{\rm s}$, because $CDF(\lambda_{\rm s},0.1)$ is small.
As $f_{\rm agn}$ increases, the PDF $P(\lambda | f_{\rm agn})$
shifts towards larger values of $\lambda$, and becomes more consistent with $\lambda_s$.
Therefore, we estimate the error of $f_{\rm agn}$ through the following steps. 
(i) We increase $f_{\rm agn}$ until we find a value which satisfies
$CDF(\lambda_{\rm s},f'_{\rm agn}) = 0.15865(0.02275)$. Any $f_{\rm agn} < f'_{\rm agn}$ should lie outside the 
$1\sigma (2\sigma)$ confidence interval (CI) of the probability distribution of $f_{\rm agn}$. 
(ii) We keep increasing $f_{\rm agn}$ until we find $CDF(\lambda_{\rm s},f''_{\rm agn}) = 0.5$. 
This $f''_{\rm agn}$ represents the median value of $f_{\rm agn}$.
(iii) We further increase  $f_{\rm agn}$ until $CDF(\lambda_{\rm s},f'''_{\rm agn}) > 1 - 0.15865(0.02275)$. 
Such $f'''_{\rm agn}$ determines the $1\sigma (2\sigma)$ upper bound of $f_{\rm agn}$.

\section{Mock Data}     \label{sec:data}

\subsection{Population models of EMRIs}    \label{sec:emri_pop}

The EMRI rate is a primary feature of the EMRI population, 
and it is mainly determined by four factors \citep{2017PhRvD..95j3012B}: 
(i) the properties of the MBH population, including the mass function and spin distribution of MBHs; 
(ii) the $M-\sigma$ relation; 
(iii) the fraction of plunges among EMRIs; (iv) the characteristic masses of
inspiralling stellar-mass compact objects.  All of these factors are uncertain
in astrophysics, and different conditions produce different EMRI population
models.  In this work, we adopt 11 EMRI population models proposed in
\cite{2017PhRvD..95j3012B}, i.e., from M1 to M12 \citep[except the M11 model,
which predicts a very low EMRI rate so that LISA and TianQin detecting almost
no EMRIs, e.g.,][]{2017PhRvD..95j3012B, 2020PhRvD.102f3016F}. We set the
eccentricities of EMRIs at plunge to be uniformly distributed in the range from
$0$ to $0.2$.  The main differences between the various models are summarized
in Table I of \cite{2017PhRvD..95j3012B}.  When simulating the EMRIs
originating in AGNs, we adjust the corresponding source parameters based on the
population models presented in \cite{2021PhRvD.104f3007P}, for instance, by
updating the redshift distribution of EMRIs and
setting all eccentricities at plunge to zero.

\subsection{Mock observations}    \label{sec:simulations}

To estimate the abilities of detectors in constraining the cosmic expansion history and 
the formation channel of EMRIs, 
we need to simulate (i) the detected EMRI catalog, (ii) the redshift probability 
distribution of the candidate host galaxies, and
(iii) the number of candidate host {\it AGNs} corresponding to each detected EMRI. 

To generate a catalog of detected EMRIs, we follow the approach of 
\cite{2017PhRvD..95j3012B} and \cite{2020PhRvD.102f3016F}. 
First, we adopt a galaxy catalog from the MultiDark cosmological simulations 
\citep{2016MNRAS.457.4340K, 2016ApJS..222...22C} 
to select the host galaxies of the EMRIs. In the selection process, 
the mass of the MBH at 
the center of the selected host galaxy matches the mass of the MBH of the corresponding EMRI. 
Second, we generate the GW signals of EMRIs using the analytic kludge Kerr (AKK) 
waveform model \citep{2004PhRvD..69h2005B, 2017PhRvD..95j3012B}. 
The reason for us to adopt this model is to facilitate comparison with 
the previous works \citep{2008PhRvD..77d3512M, 2021MNRAS.508.4512L}, 
although several new waveform models \citep{2017PhRvD..96d4005C, 
2021PhRvL.126e1102C, 2021PhRvD.103j4014H, 2022PhRvL.128w1101I} 
have been proposed recently. 
Third, we calculate the detector responses to the EMRI GW signals.

We consider three detector configurations, namely 
\begin{itemize}
 \item{\emph {TianQin}} (TQ): the default situation in which three satellites 
 form a constellation and operate in a ``3 months on + 3 months off'' mode, 
 with a mission lifetime of five years \citep{2016CQGra..33c5010L};
 \item{\emph {TianQin I+II}} (TQ I+II): twin constellations of satellites 
 with perpendicular orbital planes, that operate in a relay mode and 
 can avoid the 3 months gap in data 
 \citep{2020PhRvD.102f3016F, 2022PhRvD.105b2001L};
 \item{\emph {TianQin+LISA}} (TQ+LISA): a network composed of TianQin and LISA, 
 with the LISA configuration according to \cite{2017arXiv170200786A}. 
 In this work, we assume that LISA and TianQin have a five year mission overlap 
 and only consider EMRIs that are individually detectable by both TianQin and LISA. 
 Here we emphasize that because both LISA \citep{2024arXiv240207571C} and 
 TianQin \citep{2021PTEP.2021eA107M} are planned to be launched around 2035, 
 it is possible that there will be a window of five years of overlap. 
\end{itemize}
Finally, we set the threshold of SNR for GW detection to $20$, 
and estimate the spatial localization errors of EMRIs with 
the Fisher information matrix \citep{1994PhRvD..49.2658C, 2008PhRvD..77d2001V}. 

\begin{figure}[t]
\centering
\includegraphics[width=0.460\textwidth]{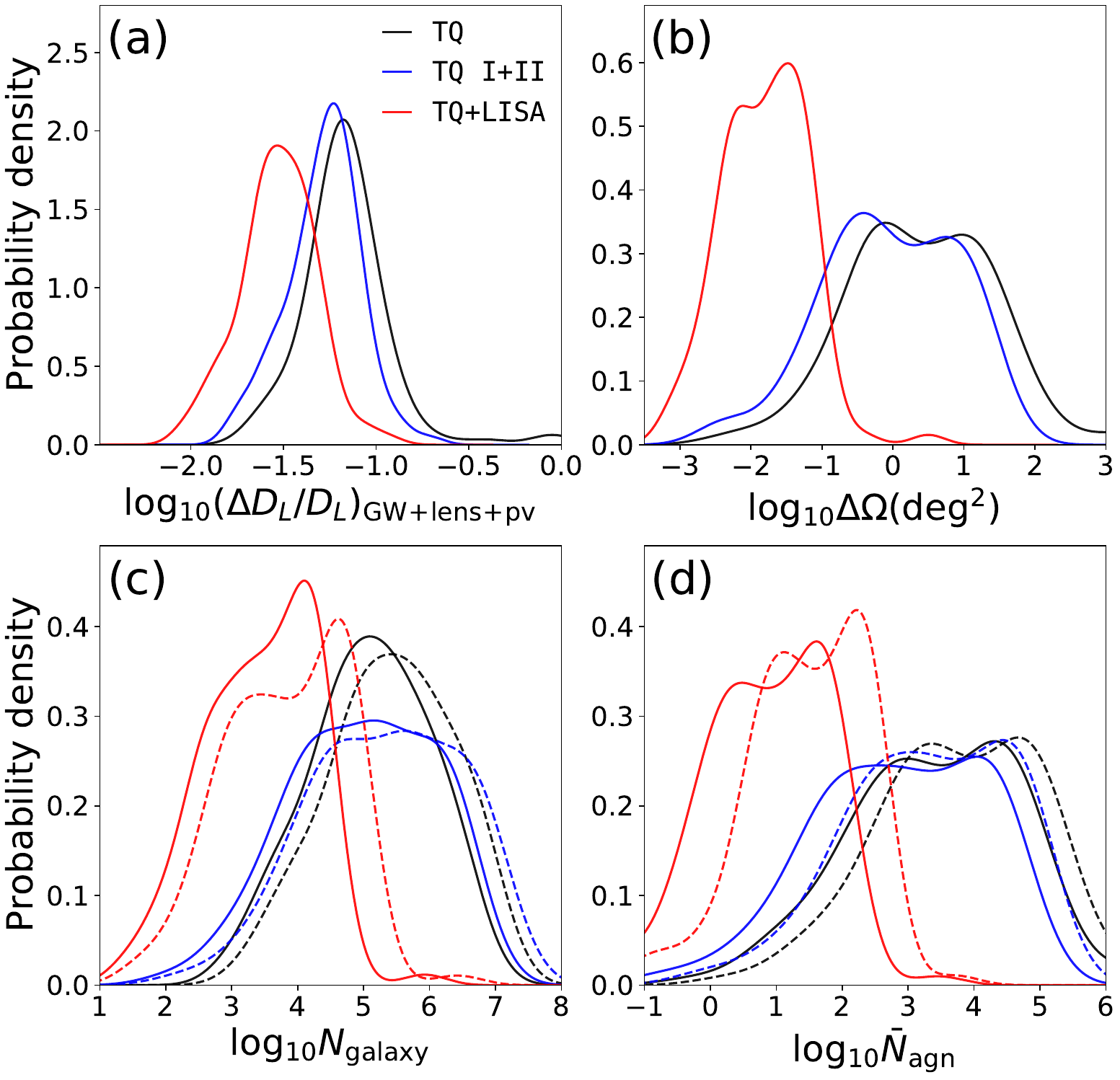}
\caption{ Panels (a), (b), (c), and (d) show, respectively, the distributions of the total $D_L$ estimation errors, 
the sky localization errors, the numbers of the candidate host galaxies, 
and the expected numbers of the candidate host AGNs. All results are 
derived based on the M1 population model. 
The colors indicate TianQin (black), TianQin I+II (blue), and TianQin+LISA (red). 
And in Panels (c) and (d), the solid and dashed lines, respectively, represent the conditions 
in which the cosmological parameters are fixed and freely varying within a prior range, 
when transforming the range of luminosity distance into the range of redshift.
}
\label{fig:dDLdOmegaNgalNagn}
\end{figure}

For example,  when using the M1 model 
\citep[one can compare our results with those in][]{2021MNRAS.508.4512L}, 
the distribution of the spatial localization errors 
(including the $D_L$ estimation and sky localization errors) of 
various detector configurations are shown in 
Panels (a) and (b) of Figure \ref{fig:dDLdOmegaNgalNagn}. 
We find that the medians of the estimation errors for luminosity distance are 
about $7\%$, $6\%$, and $3\%$ for TianQin, TianQin I+II, and TianQin+LISA, respectively, 
after accounting for additional systematic errors due to the weak lensing effects 
\citep{2010PhRvD..81l4046H, 2021MNRAS.504.3610C}
and peculiar velocities of the host galaxies \citep{2006ApJ...637...27K}.
The corresponding  medians errors for sky localization are about
$2.6 ~\!{\rm deg}^2$, $0.8 ~\!{\rm deg}^2$, and $0.02 ~\!{\rm deg}^2$, respectively. 

For the cosmological inferences, in order to simulate the redshift probability distributions of the detected EMRIs, 
we select their candidate host galaxies from 
the MultiDark galaxy catalog, and we take into account 
the selection effects of surveys. 
Here we set the limiting magnitude of the MultiDark galaxy catalog to 
$m_{\rm limit}^* = +26 ~\!{\rm mag}$ according to the survey depth 
of the Chinese Space Station Telescope \citep[CSST;][]{2019ApJ...883..203G}. 
To select the candidate host galaxies, we adopt a 
$2\sigma$ CI of the spatial localization error volumes, 
and convert the range of luminosity distance, $[D_{L;\min}, D_{L;\max}]$, 
into a range of redshift, $[z_{\min}, z_{\max}]$, using the relationship
\begin{subequations}
 \begin{align}
 D_{L;\min} &=  c(1+z_{\min}) \int_{0}^{z_{\min}} \frac{1}{H^-(z')} \D z',    \label{eq:DL_min}    \\
 D_{L;\max} &=  c(1+z_{\max}) \int_{0}^{z_{\max}} \frac{1}{H^+(z')} \D z',    \label{eq:DL_max}
 \end{align}
\end{subequations}
where $H^{-}(z)$ and $H^+(z)$ are the minimum and maximum values of $H(z)$, respectively, 
given by the cosmological prior. 
For the prior, when we focus on the constraint on $H_0$, 
we choose uniform distributions of $h \in \mathcal{U}[0.6, 0.8]$ and 
$\Omega_M \in \mathcal{U}[0.04, 0.5]$ while fixing $w_0 = -1$ and $w_a = 0$; 
when we are interested in the constraint on the dark energy EoS, 
we choose uniform distributions of $w_0 \in \mathcal{U}[-2, -0.5]$ and 
$w_a \in \mathcal{U}[-1, 1]$ while fixing $h$ and $\Omega_M$ at their default values.

The distribution of the number of the candidate host galaxies  
is shown in Panel (c) of Figure \ref{fig:dDLdOmegaNgalNagn}. 
We can see that the typical number of candidate host galaxies for an EMRI detected by 
TianQin or TianQin I+II is about $10^5 - 10^6$ after accounting for 
the uncertainties in the cosmological parameters, and that the TianQin+LISA network can reduce 
the typical number to about $10^3 - 10^4$. 
In addition, to make the mock galaxy catalog more realistic, 
we assume that all galaxies with $z > 0.6$ 
in the MultiDark galaxy catalog 
have photometric redshift (photo-$z$) errors which vary with 
the true redshifts $z_{\rm true}$ of the galaxies as
\begin{equation}   \label{eq:err_photoz} 
 \sigma_{{\rm photo-}z} (z_{\rm true}) = \left\{
 \begin{array}{ll}
 0,                        & z_{\rm true} \leq 0.6 ,    \\
 0.03(1 + z_{\rm true} ),  & z_{\rm true} > 0.6 ,
 \end{array} \right.
\end{equation}
\citep{2013ApJ...775...93D, 2021ApJ...909...53Z, 2022MNRAS.512.4593Z}. 
The reason for setting the photo-$z$ errors to zero 
for galaxies with $z \leq 0.6$ is that 
spectroscopic surveys of ground-based telescopes can cover 
this redshift range \citep{2012RAA....12.1197C, 2013AJ....145...10D, 2016arXiv161100036D}.

To constrain the formation channel of EMRIs,  
we estimate the expected number of AGNs $\lambda_i$ within the spatial localization 
error volume $\Delta V_i$ of the $i$th EMRI according to the empirical bolometric 
AGN luminosity functions $\Phi(M^*,z)$ \citep{2007ApJ...654..731H}, 
where $M^*$ represents the absolute magnitude of AGN, 
and then use this expectation to generate 
a Poisson distribution of random numbers. More quantitatively,
we calculate the expectation $\lambda_i$ with 
\begin{equation}  \label{eq:nAGN_expected}
\lambda_i = \Delta V_i \int_{-\infty}^{M_{{\rm limit},i}^{*}}    \Phi(M^*,z) \D M^* ,
\end{equation}
where 
\begin{align}  
M_{{\rm limit},i}^{*} \approx  m_{\rm limit}^{*{\rm agn}} - 25 - 5 \lg \!\left(\! \frac{D_{L,i}}{1 \!~{\rm Mpc}} \!\right)\! - BC
\nonumber 
\end{align}
represents the observational limiting absolute magnitude for the candidate host AGN
of the $i$th EMRI at a distance of $D_{L,i}$, and $BC$ is a factor for bolometric correction \citep{2020A&A...636A..73D}. 
In this work, we set the limiting magnitude of the AGN survey 
to $m_{\rm limit}^{*{\rm agn}} = +24.4 ~\!{\rm mag}$ according to 
the depth in the shallowest band of the
CSST survey as well as taking into account the need for the 
spectroscopic identifications of AGNs \citep{2019ApJ...883..203G}. 
Taking the M1 model as an example, the distributions of the numbers 
of candidate host AGNs of the EMRIs for various detector configurations 
are shown in Panel (d) of Figure \ref{fig:dDLdOmegaNgalNagn}. 
We can see that for the same EMRI, the number of candidate host AGNs is about $1\%$ 
of the number of the normal galaxies which are candidate hosts.  
This fraction is consistent with the AGN fraction among all galaxies.

\section{Constraints on cosmological parameters}     \label{sec:result_cosmo}

In this section, we use the Bayesian framework described in Section \ref{sec:Bayesian_framework} 
to calculate the posterior probability distributions of various cosmological parameters, 
including the standard $\Lambda$CDM model parameters $(H_0, \Omega_M)$ and 
the dark energy EoS parameters $(w_0, w_a)$. 
The essence of the computational process is fitting 
the $D_L - z$ relation with Markov Chain Monte-Carlo (MCMC) sampling. 
In the following, we adopt a total of 11 population models, 
from M1 to M12 except M11, 
to evaluate the capability of future EMRI observations 
in constraining the cosmological parameters.

\subsection{Constraints on $H_0$ and $\Omega_M$}    \label{sec:result_cosmo_h0}

\begin{figure*}[ht]
\centering
\includegraphics[width=0.80\textwidth]{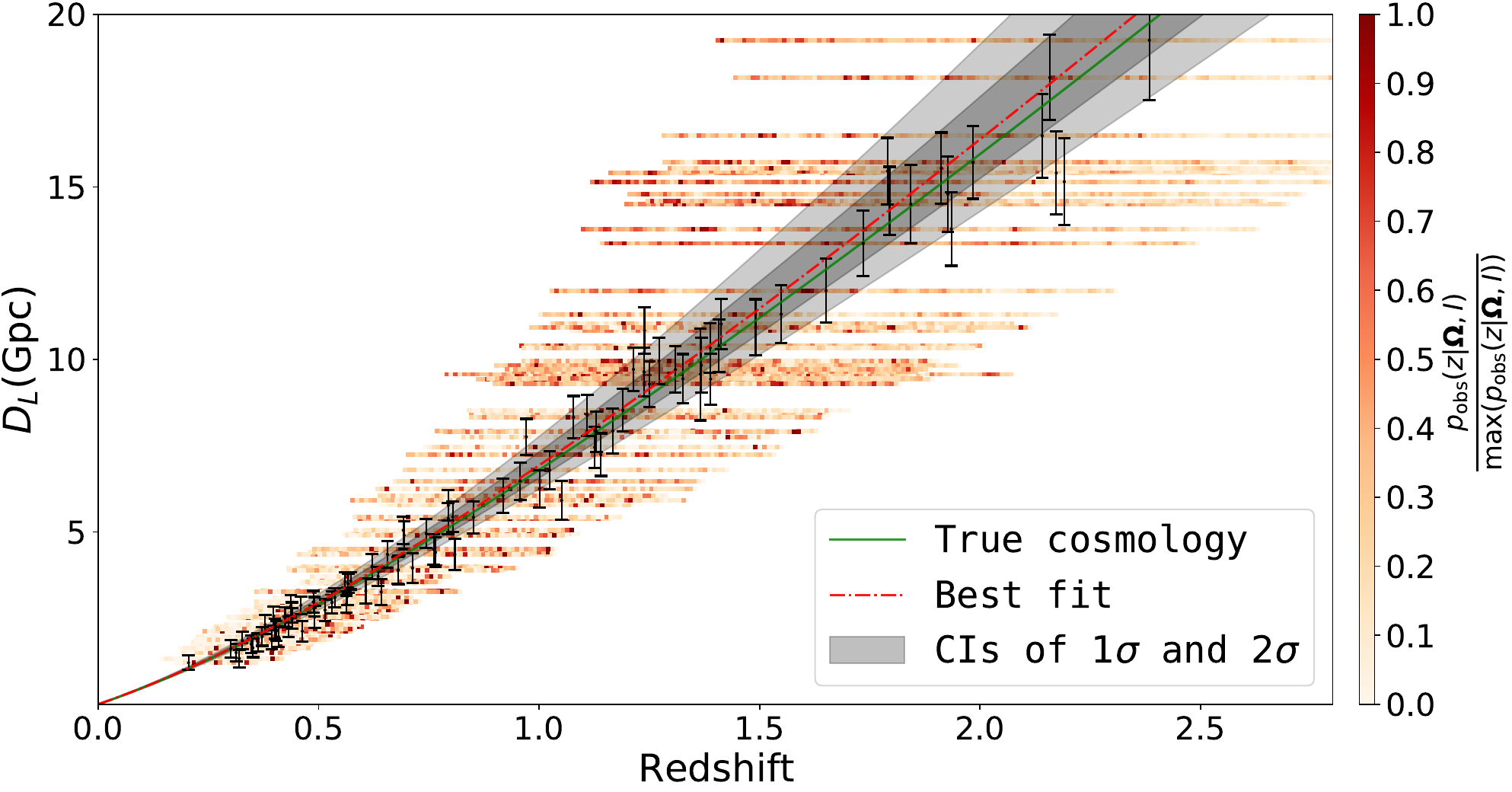}
\caption{Typical example of fitting the $D_L - z$ relation 
which is derived from the EMRI dark sirens detectable by the TianQin+LISA network. Here we use the fiducial EMRI population model. 
The horizontal colored serial squares represent the probability distribution of redshift (normalized by the maximum) 
derived from the observed candidate host galaxies, 
and the color hues from white to red show the probability of redshift.
The luminosity distances shown here represent the mean values,
and their errors ($1\sigma$ CI) are indicated by the black error bars.   
The solid green line represents the true cosmological model adopted by our simulations.
The dot-dashed red line represents the best-fit cosmological model, 
while the two shaded regions represent the $1 \sigma$ and $2\sigma$ CIs. 
}
\label{fig:fitting_DLz}
\end{figure*}

\begin{figure}[htbp]
\centering
\includegraphics[width=0.460\textwidth]{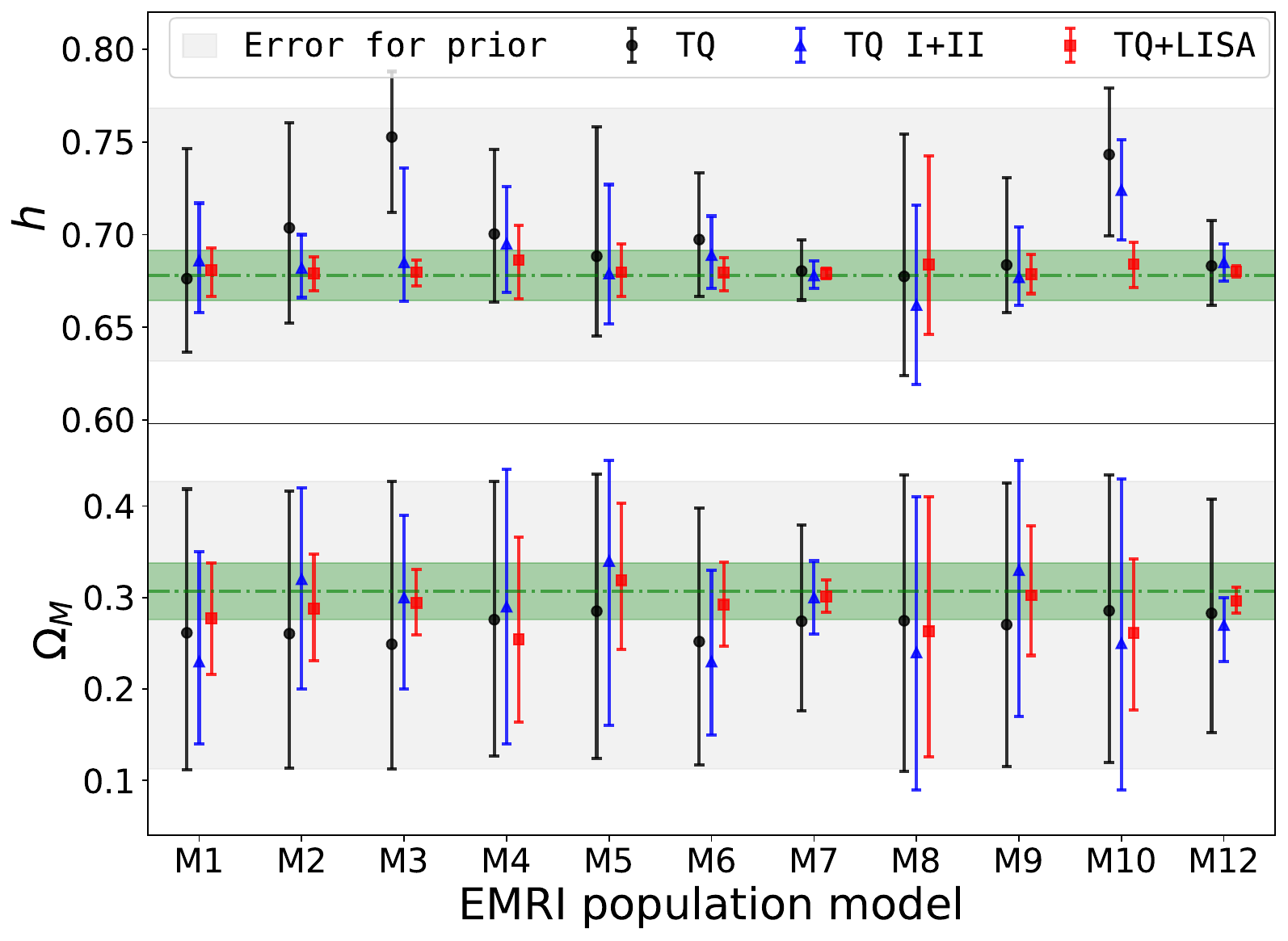}
\caption{Constraints on $h$ and $\Omega_M$ given by the EMRIs detectable by
TianQin (black), TianQin I+II (blue), and TianQin+LISA (red). 
The horizontal axis represents the different EMRI population models. 
The error bars correspond to an $1\sigma$ CI. 
The gray-shaded area represents the $1\sigma$ equivalent error of the prior, 
and the green-shaded region represents an error scale of $2\%$ ($10\%$) for $h$ ($\Omega_M$). 
}
\label{fig:constraint_H0M}
\end{figure}

We first fit the $D_L - z$ relation, using the
luminosity distances from EMRI observations and 
the redshifts extracted from galaxy surveys. The result is shown in
Figure \ref{fig:fitting_DLz}, and the 
corresponding constraints on $H_0$ and $\Omega_M$ 
are shown in Figure \ref{fig:constraint_H0M}. Note that 
in this section we fix the parameters of the dark energy EoS 
to $w_0 = -1$ and $w_a = 0$.

For the Hubble-Lema\^itre constant $H_0$, 
under the fiducial population model, i.e., the M1 model, the expected error 
is about $8.1\%$ for TianQin, and 
TianQin I+II can reduce this error to about $4.4\%$. 
Under the other population models from M2 to M12, the error of $H_0$ 
varies considerably from model to model.
The variations are mainly caused by the fact that the detection rates of EMRIs predicted by 
various models vary considerably \citep[see][]{2017PhRvD..95j3012B, 2020PhRvD.102f3016F}. 
In general, TianQin  
is able to achieve a precision in the range of about $3.4\%-9.6\%$, 
while that for  TianQin I+II is expected to be in the range of about $1.5\% - 7.2\%$. 

For the fractional matter density parameter $\Omega_M$,  
in almost all population models the TianQin configuration cannot
provide reasonable constraint  
(comparing the error scales corresponding to the priors and the posteriors), 
except for the M6, M7, and M12 models. 
But TianQin I+II can significantly improve the constraint on $\Omega_M$. 
More quantitatively, under the M6, M7, and M12 models, 
TianQin is likely to constrain $\Omega_M$ 
to a precision of about $46\%$, $33\%$, and $42\%$, respectively, 
whereas TianQin I+II would be able to constrain $\Omega_M$ to 
a precision of about $29\%$, $13\%$, and $11\%$, respectively.

If TianQin and LISA could form a network, i.e., TianQin+LISA, the constraints
on $H_0$ and $\Omega_M$ are significantly improved in all population models, as
is shown by the red error bars in Figure \ref{fig:constraint_H0M}.
For example, in the M1 model, the errors of $H_0$ and $\Omega_M$ can
shrink to about $1.9\%$ and $20\%$, respectively. Under the other models, the
precision of $H_0$ can be improved to about $0.4\% - 7.1\%$, and the error of
$\Omega_M$ would reduce to the range of about $13\% - 43\%$.  For completeness, we
summarize the expected precisions on $H_0$ and $\Omega_M$ for various cases in
Table \ref{tab:constraint_errs}.

In addition, the importance of improving the sky localization by employing the
TianQin+LISA network can be clearly seen in the upper panel of Figure~\ref{fig:constraint_H0M}.
In particular, under Models  M3 and M10, the posterior of $h$ provided by a single detector, such as
TianQin or TianQin I+II, could be significantly biased with respect to the true value. The
bias is related to the paucity of EMRIs that can be precisely localized
[see Panel (b) of Figure~\ref{fig:dDLdOmegaNgalNagn}], which increases the fluctuation in the
posterior distribution of the redshift of an EMRI. Using TianQin+LISA, however, more EMRIs 
can be precisely localized, and the precision reduces the fluctuation of the redshift distribution
function for each EMRI.  We notice that a similar trend can be seen in earlier studies
\citep{2021MNRAS.508.4512L, 2022PhRvD.105d3509M,
2022PhRvR...4a3247Z,2022SCPMA..6559811Z}.

\subsection{Constraints on $w_0$ and $w_a$}    \label{sec:result_cosmo_w0wa}

\begin{figure}[htbp]
\centering
\includegraphics[width=0.460\textwidth]{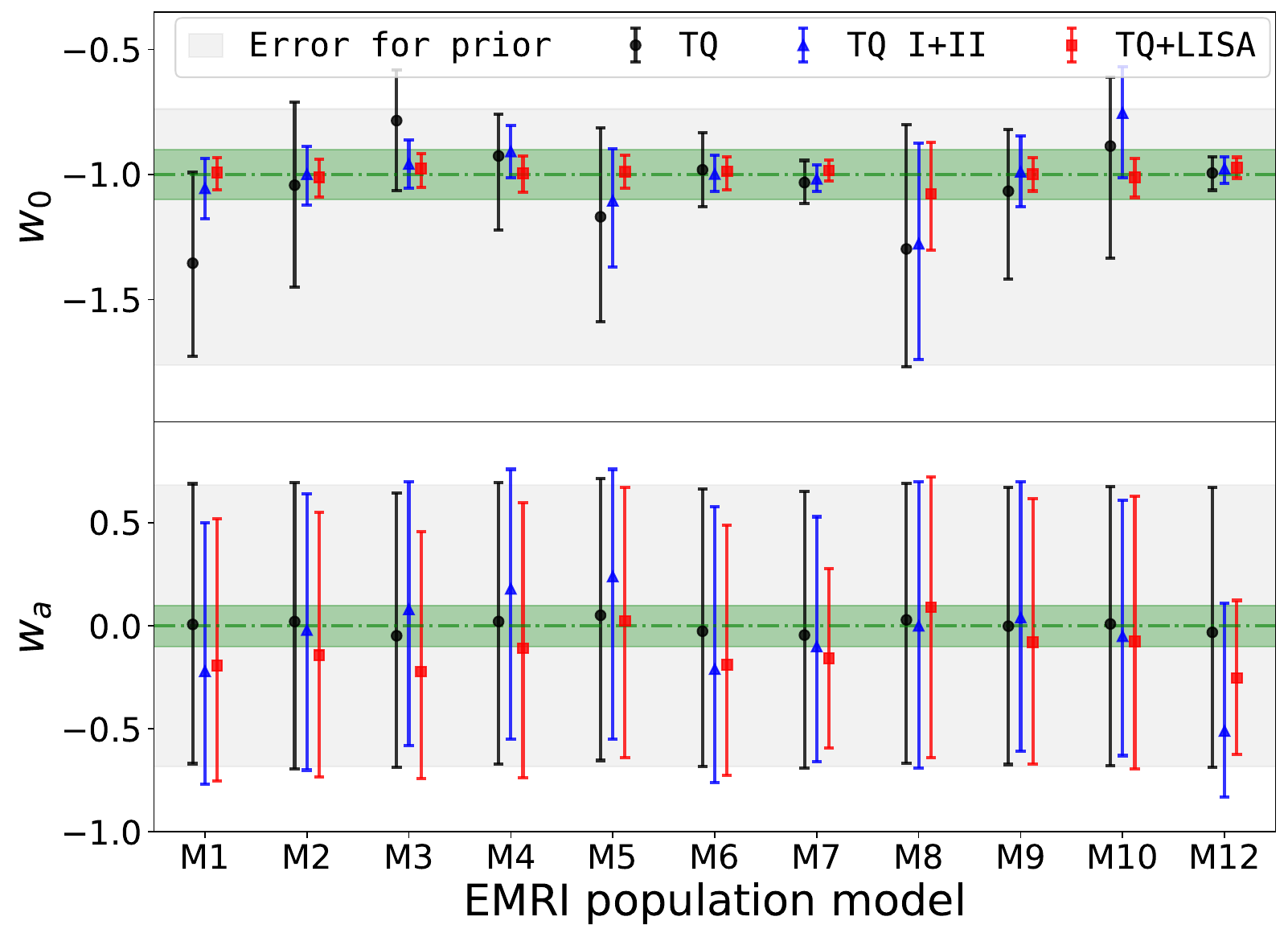}
\caption{Same as Figure \ref{fig:constraint_H0M}, but for $w_0$ and $w_a$. 
The green shades represent error scales of $\Delta w_0 = 0.1$ and $\Delta w_a = 0.1$. 
}
\label{fig:constraint_w0wa}
\end{figure}

The errors on the dark energy EoS parameters $w_0$ and $w_a$ 
are shown in Figure \ref{fig:constraint_w0wa}. 
Here we fix $h = 0.678$ and $\Omega_M = 0.307$. 
We find that only $w_0$ can be effectively constrained 
and $w_a$ is not constrained in almost all population models. 
Even under the M7 and M12 models, 
despite the fact that they have optimistic EMRI rates such that
TianQin I+II and TianQin+LISA can detect a total number of
more than one thousand EMRIs, still, a stringent constraint on $w_a$ 
cannot be obtained. 
The much better constraint on $w_0$ than on $w_a$ implies that 
reconstructing
the evolution of the dark energy with redshift requires
much better EMRIs than those needed to
infer the deviation of the dark-energy EoS with 
respect to the value $w = -1$ of the standard model.

More specifically, under the fiducial population model, the error of $w_0$ 
can reach about $37\%$ for TianQin, about $12\%$ for TianQin I+II, 
and about $6.5\%$ for TianQin+LISA. 
Under the other population models, i.e., from M2 to M12 except M11, 
the expected errors of $w_0$ can reach 
$6.7\% - 48\%$ for TianQin, about $5.4\% - 43\%$ for TianQin I+II, 
and about $4.2\% - 22\%$ for TianQin+LISA. 
Similar to the constraints on $H_0$, the errors on $w_0$ 
also vary considerably across population models.
Finally, in Table \ref{tab:constraint_errs} we also summarize the expected  
precisions in the constraints on $w_0$ and $w_a$ for various models.

\begin{table*}[ht]
  \caption{Expected relative errors of the  $\Lambda$CDM cosmological model parameters $(H_0, \Omega_M)$ and the CPL dark energy model parameters $(w_0, w_a)$ for the 11 EMRI 
  population models and assuming three different detector 
  configurations: TianQin, TianQin I+II and TianQin+LISA. The bar ``$-$'' 
	means no effective constraint.}
    \vspace{-6pt}
    \renewcommand\arraystretch{1.1}
    \centering
    \begin{tabular}{c|cc|cc|cc|cc|cc|cc}
        \hline
        \hline
        EMRI & \multicolumn{12}{c}{Expected relative error}  \\
        \cline{2-13}
        population & \multicolumn{4}{c|}{TianQin} & \multicolumn{4}{c|}{TianQin I+II} & \multicolumn{4}{c}{TianQin+LISA}  \\
        \cline{2-13}
        model         & $\frac{\Delta H_0}{H_0}$  & $\frac{\Delta \Omega_M}{\Omega_M}$   & $|\frac{\Delta w_0}{w_0}|$ & $\frac{\Delta w_a}{1} $   & $\frac{\Delta H_0}{H_0}$  & $\frac{\Delta \Omega_M}{\Omega_M}$      & $|\frac{\Delta w_0}{w_0}|$ & $\frac{\Delta w_a}{1} $   & $\frac{\Delta H_0}{H_0}$  & $\frac{\Delta \Omega_M}{\Omega_M}$    & $|\frac{\Delta w_0}{w_0}|$ & $\frac{\Delta w_a}{1} $  \\
        \hline
        M1     & 8.1\% & $-$   & 37\% & $-$   & 4.4\% & 34\%   & 12\%  & $-$   & 1.9\% & 20\%   & 6.5\% & $-$ \\
        M2     & 7.9\% & $-$   & 37\% & $-$   & 2.5\% & 36\%   & 12\%  & $-$   & 1.4\% & 19\%   & 7.5\% & $-$ \\
        M3     & 5.6\% & $-$   & 24\% & $-$   & 5.3\% & 31\%   & 9.6\% & $-$   & 1.0\% & 12\%   & 6.7\% & $-$ \\
        M4     & 6.1\% & $-$   & 23\% & $-$   & 4.2\% & $-$    & 10\%  & $-$   & 2.9\% & 33\%   & 7.2\% & $-$ \\
        M5     & 8.3\% & $-$   & 39\% & $-$   & 5.5\% & 47\%   & 24\%  & $-$   & 2.1\% & 26\%   & 6.6\% & $-$ \\
        M6     & 4.9\% &46\%   & 15\% & $-$   & 2.9\% & 29\%   & 7.3\% & $-$   & 1.3\% & 15\%   & 6.5\% & $-$ \\
        M7     & 2.4\% &33\%  & 8.7\% & $-$   & 1.1\%  & 13\%  & 5.4\% & $-$   & 0.4\% & 5.8\%  & 4.1\% & 44\% \\
        M8     & 9.6\% & $-$   & 48\% & $-$   & 7.2\%  & $-$    & 43\% & $-$   & 7.1\% & 46\%   & 22\%  & $-$ \\
        M9     & 5.4\% & $-$   & 30\% & $-$   & 3.1\%  & 46\%   & 14\% & $-$   & 1.6\% & 23\%   & 6.6\% & $-$ \\
        M10    & 5.9\% & $-$   & 36\% & $-$   & 4.0\%  & $-$    & 22\% & $-$   & 1.8\% & 27\%   & 7.8\% & $-$ \\
        M12    & 3.4\% &42\%  & 6.7\% & $-$   & 1.5\%  & 11\%  &5.4\%  &47\%   & 0.4\% & 4.7\%  & 4.2\% & 37\% \\
        \hline
        \hline
    \end{tabular}
    \label{tab:constraint_errs}
\end{table*}

\section{Constraining the formation channel of EMRIs}     \label{sec:result_astrop}

\begin{figure}[htbp]
\centering
\includegraphics[width=0.460\textwidth]{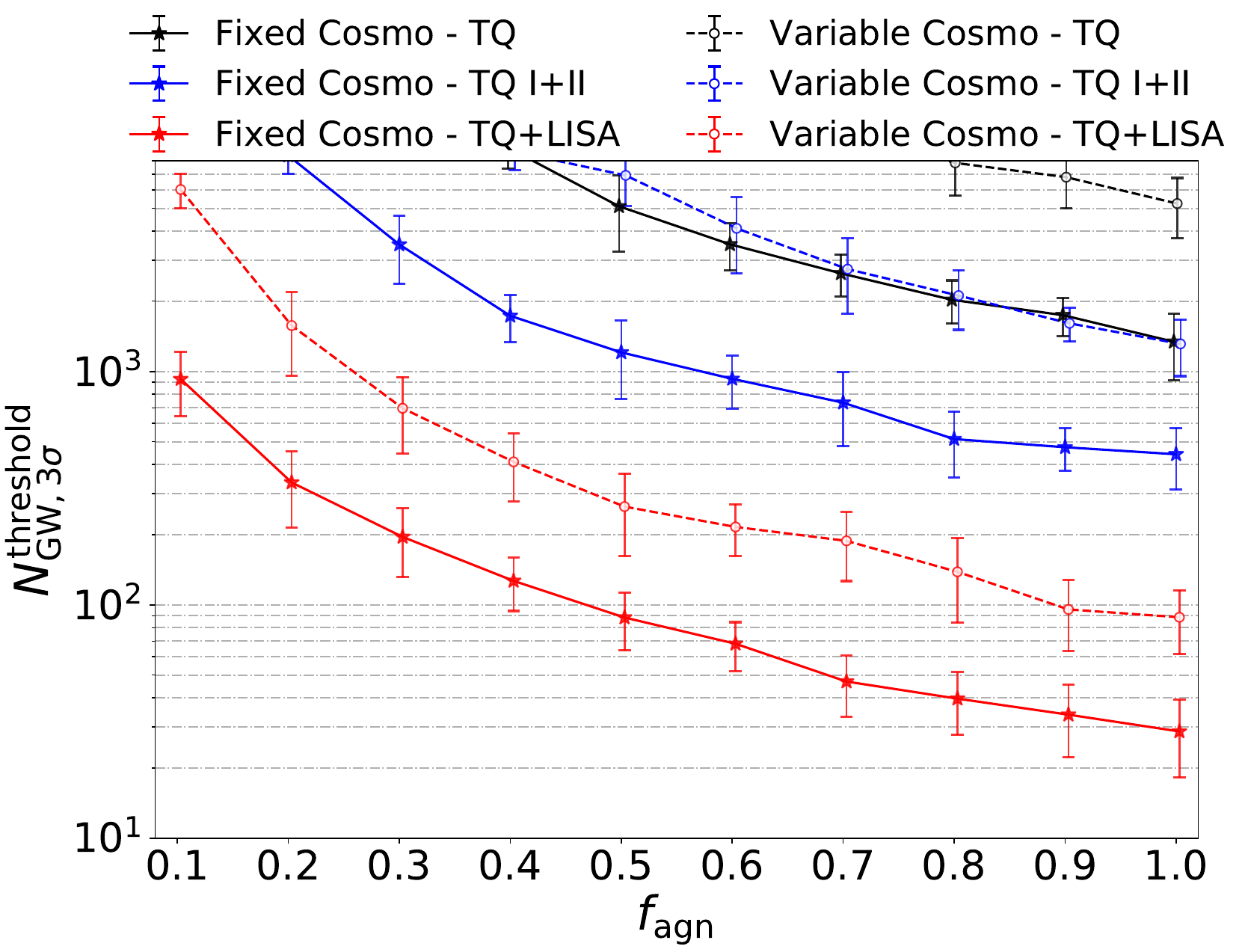}
\caption{Minimum number $N_{{\rm GW},3\sigma}^{\rm threshold}$ of detected EMRI events 
required to testify the EMRI-AGN correlation with a $3\sigma$ significance. 
The black, blue, and red error bars and lines 
correspond to 
TianQin, TianQin I+II, and TianQin+LISA observations. 
The solid lines 
represent the case in which both $H_0$ and $\Omega_M$ are fix to the true values, 
and the dashed lines with corresponding error bars represent the case in which 
both $H_0$ and $\Omega_M$ are freely variable within the prior.
The error bars correspond to $1\sigma$ CI. 
}
\label{fig:nGW_fAGN}
\end{figure}

So far, we have assumed that each candidate host galaxy has the same probability 
of hosting an EMRI, be it a normal galaxy or an AGN. 
Now we drop this assumption and
use the statistical framework described in Section \ref{sec:Poisson_framework} 
to constrain $f_{\rm agn}$, which is the fraction of EMRIs which reside in AGNs. 
Since the statistical significance increases with the number of GW sources, 
we denote $N_{{\rm GW},3\sigma}^{\rm threshold}$ as the minimum number 
of detected EMRIs that is required to reject the null hypothesis at 
a significance level of $3\sigma$, and we study its dependence on 
the value of $f_{\rm agn}$.

Here, we follow the works of \citet{2017PhRvD..95j3012B} and 
\citet{2021PhRvD.104f3007P} 
to generate the MBH mass function and its evolution with redshift.
Correspondingly, 
the dimensionless MBH spin is set to $0.98$ and the mass of the
stellar BH is fixed at  $10 ~\! M_{\odot}$.
The above choices are commonly used in the previous works 
\citep{2012MNRAS.423.2533B, 2013MNRAS.429.3155A, 2019ApJ...882L..24A,2023PhRvX..13a1048A}, 
thus we adopt them for easier comparison with the previous works.
The eccentricities of EMRIs are expected to be significantly different in different models.
For the EMRIs in normal galaxies, we assumed a uniform distribution of 
the eccentricities at plunge within the range of $[0, 0.2]$, as is proposed by 
\cite{2017PhRvD..95j3012B}; 
for the EMRIs originating in AGNs, we set their eccentricities to zero, 
to account for the damping of eccentricity suggested by \cite{2021PhRvD.104f3007P}.
When we convert the error of luminosity distance to the redshift error 
using Equations (\ref{eq:DL_min}) and (\ref{eq:DL_max}), 
we assume two cosmological conditions, namely: 
(i) ``fixed cosmology'' in which 
the cosmological parameters are fixed to their true values, 
and (ii) ``variable cosmology'' where 
the cosmological parameters $H_0$ and $\Omega_M$ are 
freely variable within their priors. 
We consider the latter condition to account for 
the tension in the current measurements of the Hubble constant 
\citep{2017NatAs...1E.121F, 2020A&A...641A...6P, 2021ApJ...908L...6R, 
2021CQGra..38o3001D, 2022NewAR..9501659P}. 
The resulting dependence of the required minimum number 
$N_{{\rm GW},3\sigma}^{\rm threshold}$ 
on the fraction $f_{\rm agn}$ 
is shown in Figure \ref{fig:nGW_fAGN}. 
We can see that $N_{{\rm GW},3\sigma}^{\rm threshold}$
decreases with increasing $f_{\rm agn}$ in all cases. 

In the ``fixed cosmology'' condition, TianQin would need to detect 
at least $1300$ EMRIs to be able to reject the null hypothesis 
and establish the EMRI-AGN correlation if $f_{\rm agn} = 1$. 
Under the same condition, TianQin I+II needs to detect 
at least about $450$ EMRIs and TianQin+LISA needs to detect 
at least about $30$ EMRIs. 
In the ``variable cosmology'' condition, if $f_{\rm agn} = 1$ 
the minimum number $N_{{\rm GW},3\sigma}^{\rm threshold}$ required for TianQin, 
TianQin I+II, and TianQin+LISA are about $5000$, $1300$, and $90$, respectively. 
The increase of $N_{{\rm GW},3\sigma}^{\rm threshold}$ in the case of variable cosmology 
is caused by an increase of the redshift range when transforming 
the luminosity distance range into the redshift space. 
For different cases, $N_{{\rm GW},3\sigma}^{\rm threshold}$ 
as a function of $f_{\rm agn}$ can be approximate as 
$N_{{\rm GW},3\sigma}^{\rm threshold} \propto f_{\rm agn}^{-1.6}$.

\begin{figure}[tbp]
\centering
\includegraphics[width=0.460\textwidth]{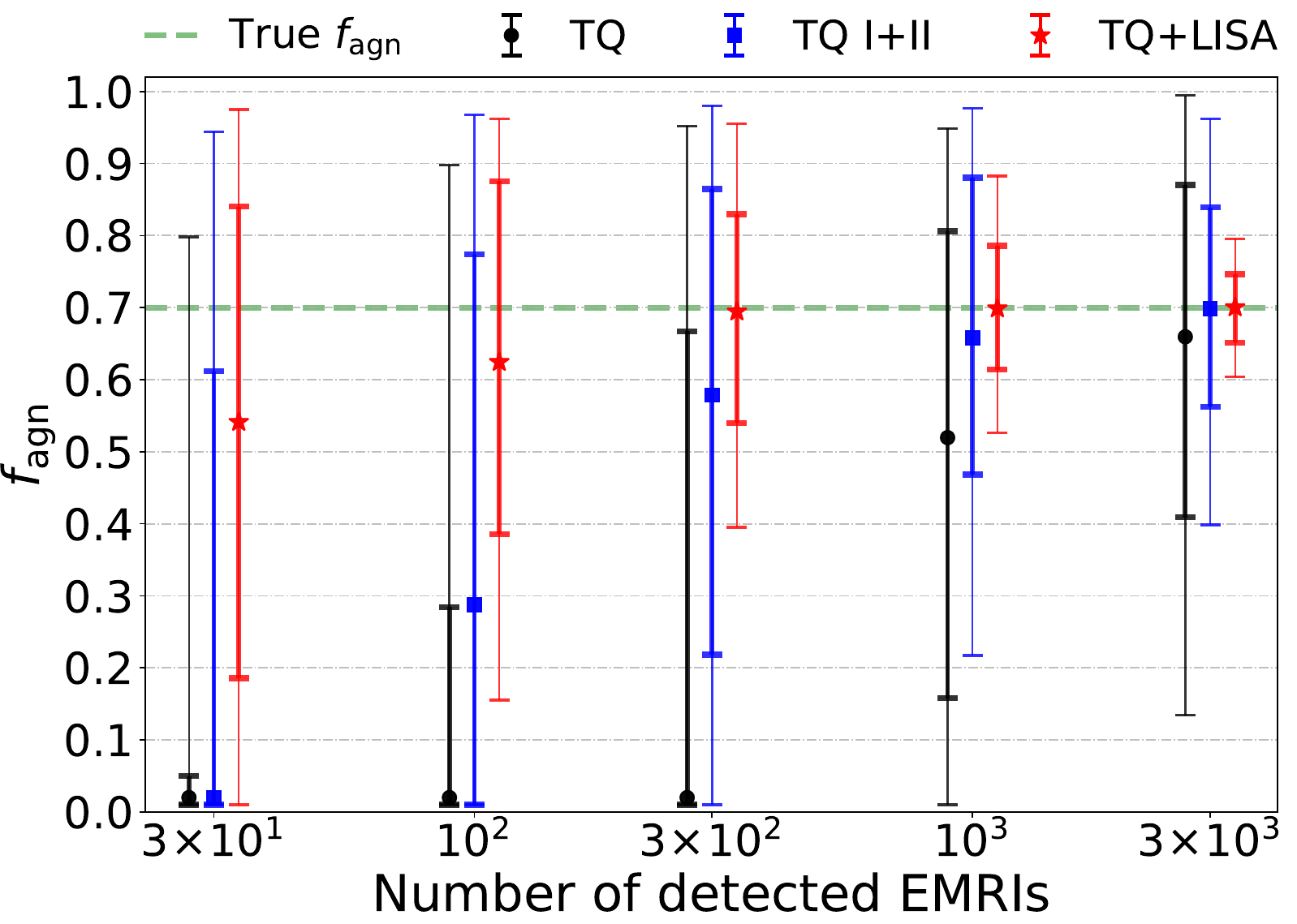}
\caption{
The error in constraining $f_{\rm agn}$ as a function of the number of detected EMRIs.  
Here we fix the cosmological parameters to their true values. 
The black, blue, and red error bars correspond to 
TianQin, TianQin I+II, and TianQin+LISA. 
The thick and thin error bars correspond to $1\sigma$ and $2\sigma$ CIs, respectively. 
The dots in the middle of the error bars represent the expectations. 
The green dashed horizontal line represents the injected $f_{\rm agn}$ in the simulations. 
}
\label{fig:fAGNerr_nGW}
\end{figure}

After testing the EMRI-AGN correlation, we proceed to estimate the value
of $f_{\rm agn}$ using Equation (\ref{eq:CDF-fagn}) and Monte-Carlo
simulations.  As an illustration, we set the fraction of EMRIs originating in
AGNs to $f_{\rm agn}=0.7$. The variation of the error of $f_{\rm agn}$ with the
number of detected EMRIs for different detector configurations is shown in
Figure~\ref{fig:fAGNerr_nGW}.  We find that as the number of detected EMRIs
approaches or exceeds the threshold number $N_{{\rm GW},3\sigma}^{\rm
threshold}$ (refer to Figure \ref{fig:nGW_fAGN}), the constraint on $f_{\rm
agn}$ converges toward the injected value, with the constraining error
diminishing as the number of detected EMRIs increases.  When the number of
detected EMRIs is considerably lower than $N_{{\rm GW},3\sigma}^{\rm
threshold}$, the constraint on $f_{\rm agn}$ is consistent with $f_{\rm agn} =
0.01$ (the fraction of AGNs among all galaxies we assumed), but with
large error.  
Furthermore, for an equivalent number of EMRI detections, the TianQin+LISA
network significantly enhances the precision of constraining $f_{\rm agn}$. For
instance, with 1000 EMRI detections, the errors corresponding to $1\sigma$ CIs
are $\Delta f_{\rm agn} \sim 0.33$ for TianQin and $\Delta f_{\rm agn} \sim
0.21$ for TianQin I+II, whereas the error for TianQin+LISA shrinks to $\Delta
f_{\rm agn} \sim 0.09$.  

We would like to note that the real EMRI population in the universe is
more completed than the modeled ones presented in this work.  For examples,
both the spins of the MBHs \citep{2014ApJ...794..104S, 2019MNRAS.490.4133B,
2022MNRAS.514.2568S} and the masses of the stellar-mass BHss
\citep{2022PhRvD.105h3005P, 2023PhRvX..13a1048A} should follow certain
distributions, but we have adopted fixed values for them.  In addition, the
EMRIs originating in normal galaxies  \citep{2007MNRAS.378..129H,
2008ApJ...675..604Y, 2013MNRAS.435.3521B, 2019PhRvD..99l3025A,
2020MNRAS.498L..61H, 2022PhRvD.106l4028F} or in those post-merger galaxies
hosting MBHBs \citep{2022MNRAS.516.1959M, 2022ApJ...927L..18N,
2023ApJ...955L..27N} may exhibit eccentricities much higher than $0.2$, and in
AGNs the stellar-mass BHs would interact with the accretion disks
\citep{2011PhRvL.107q1103Y, 2023MNRAS.526.5612C}.  All these factors could
affect the duration, GW amplitude, or extra modes of an EMRI, and consequently
affect its detectability.  Nevertheless, our work serves as a key step towards
testing the EMRI-AGN correlation.

\section{Discussions}    \label{sec:discussion}

\subsection{Advantage of combining EMRIs with AGN observation}    \label{sec:discuss_EMRIplusAGNcatalog}

Once either of the following two conditions is met, we can use AGN catalogs to
extract the redshift information of a detected EMRI (or at least assign higher
weights to the AGNs): (i) the fraction of the EMRIs originate in AGNs, $f_{\rm
agn}$, is statistically constrained to be close to $1$ by our method; and (ii)
the detected EMRI has a zero eccentricity at the time of plunge \citep[see,
e.g., ][for explanation]{2021PhRvD.103j3018P, 2021PhRvD.104f3007P,
2023LRR....26....2A}.  We note that LISA and TianQin can measure the
eccentricities of EMRIs to a precision of $\Delta e \approx 10^{-6}$
\citep{2017PhRvD..95j3012B, 2020PhRvD.102f3016F}, so they can differentiate the
formation channels. 

\begin{figure}[tbp]
\centering
\includegraphics[width=0.230\textwidth]{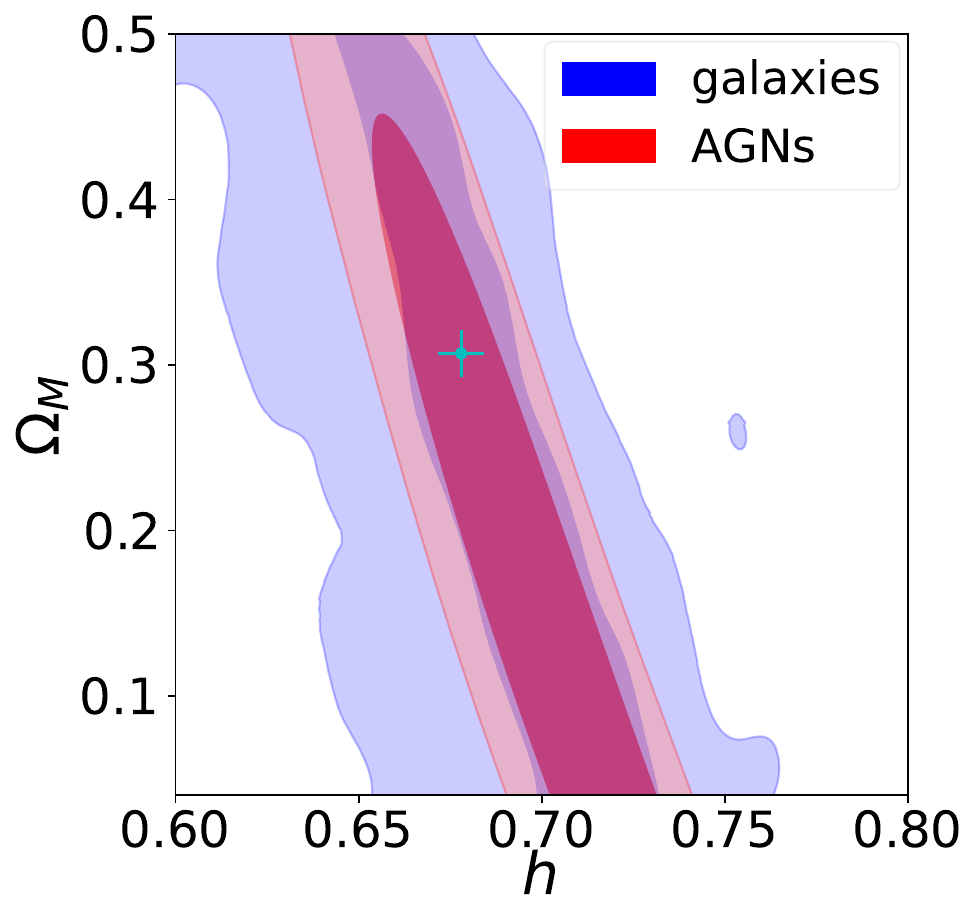}
\includegraphics[width=0.2350\textwidth]{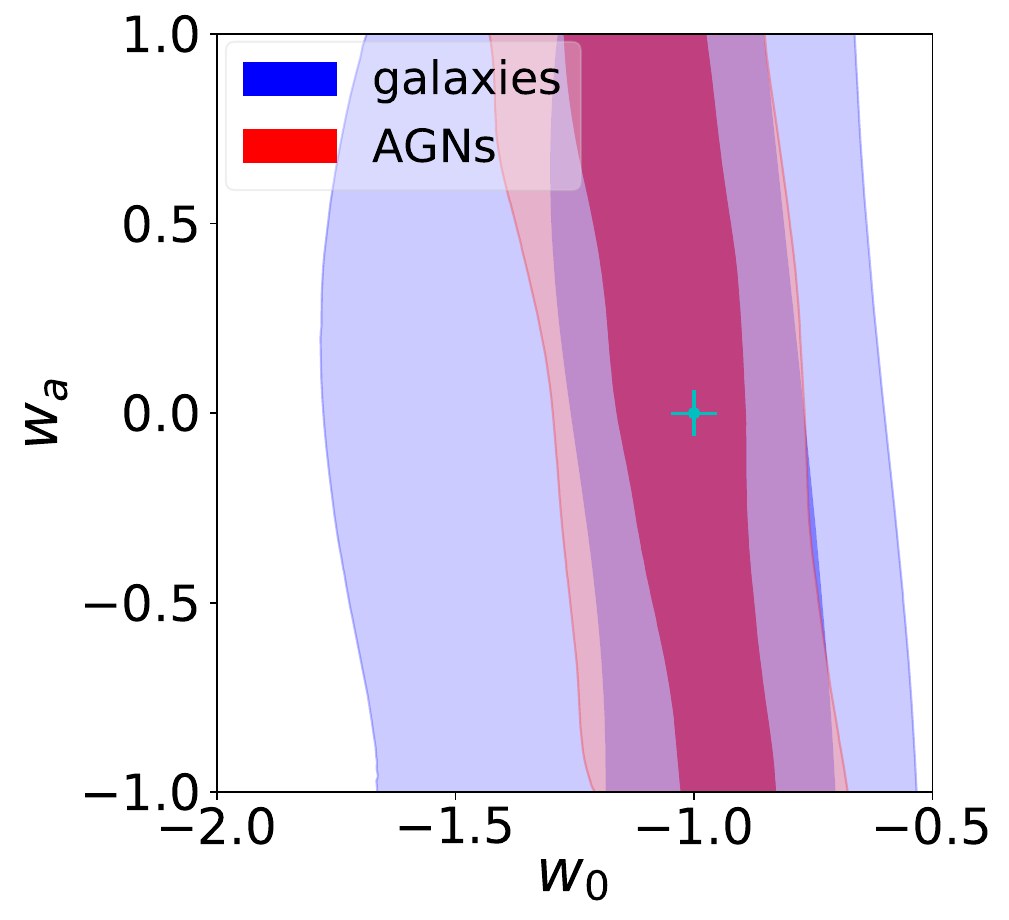}
\caption{Constraints on $(h, \Omega_M)$ (left panel) and 
$(w_0, w_a)$ (right panel) by combining the EMRIs detected by TianQin with AGN catalogs. 
The eccentricities at plunges are set to zero, and the other 
parameters are derived from the M1 model. 
The blue contours represent the constraints derived 
from candidate host galaxies 
and the red contours represent the constraints from candidate host AGNs. The 
two levels of contours correspond to $1\sigma$ and $2\sigma$ CIs, respectively. 
The cyan crosses mark the injected cosmological parameters in our simulations. 
}
\label{fig:H0Om_w0wa_AGNconstraints}
\end{figure}

Assuming that all EMRIs come from AGNs, and using only the AGN catalog
to look for the candidate hosts for EMRIs, we repeat the simulation shown in 
Section \ref{sec:simulations} and derive the posteriors for the cosmological parameters.
More specifically, we consider the EMRIs detected by TianQin in the M1 population model 
as an example. 
The AGN catalog is based on the MultiDark catalog but we use the limiting
magnitude of CSST to select observable AGNs. 
The result is shown in Figure \ref{fig:H0Om_w0wa_AGNconstraints}. 
We find that the constraints on cosmological parameters are improved.
More quantitatively, 
by switching from the galaxy catalog to the AGN catalog, 
the precision of $H_0$ improves from about $3.7\%$ to about $3.2\%$, and 
the precision of $w_0$ improves from about $25\%$ to about $13\%$. 
For a fair comparison, here we do not consider the redshift errors 
in either candidate host galaxies or AGNs.

These improvements mainly benefit from the fact that 
AGNs constitute only about $(1-10)\%$ of all galaxies
\citep[e.g., compare the galaxy and AGN luminosity functions in][]{2005ApJ...631..126D, 2007ApJ...654..731H}. 
When $f_{\rm agn}$ is close to $1$ or the detected EMRIs have 
zero eccentricities at plunge, we can derive more precisely 
the redshifts of EMRIs with the help of AGN catalogs.
These results also imply that circular EMRIs are better 
than those eccentric ones in constraining cosmological parameters. 

In addition, we would like to point out that EMRIs may be 
accompanied by EM counterparts 
\citep{2019ApJ...886L..22W, 2022ApJ...933..225W,2023ApJ...956L...2L, 2023LRR....26....2A}. 
The EM signals could be detected by the
current and planned time-domain survey telescopes, such as 
the Wide Field Survey Telescope (WFST) \citep{2023SCPMA..6609512W, 2024arXiv240301686H}, 
the Multi-channel Photometric Survey Telescope (Mephisto) \citep{2020SPIE11445E..7MY}, 
Vera Rubin Observatory \citep[also known as LSST;][]{2019ApJ...873..111I},
the Einstein Probe \citep{2015arXiv150607735Y}, 
Athena \citep{2013arXiv1306.2307N, 2020NatAs...4...26M}, 
the Five-Hundred Aperture Spherical Radio Telescope (FAST) \citep{2011IJMPD..20..989N},
or the Square Kilometre Array (SKA) \citep{2015aska.confE..37J}. 
Detecting an EM counterpart could help us narrow down 
the candidates host galaxies, or even uniquely identify the host galaxy of an EMRI.
Such detections will enable us to
not only improve the precisions of the cosmological parameters 
due to the more accurate redshift information of the EMRIs, 
but also better understand the formation channel of EMRIs. 

\subsection{Effects of eccentricity on the 
spatial localization of EMRIs}    \label{sec:discuss_e0plunge}

\begin{figure}[t]
\centering
\includegraphics[width=0.40\textwidth]{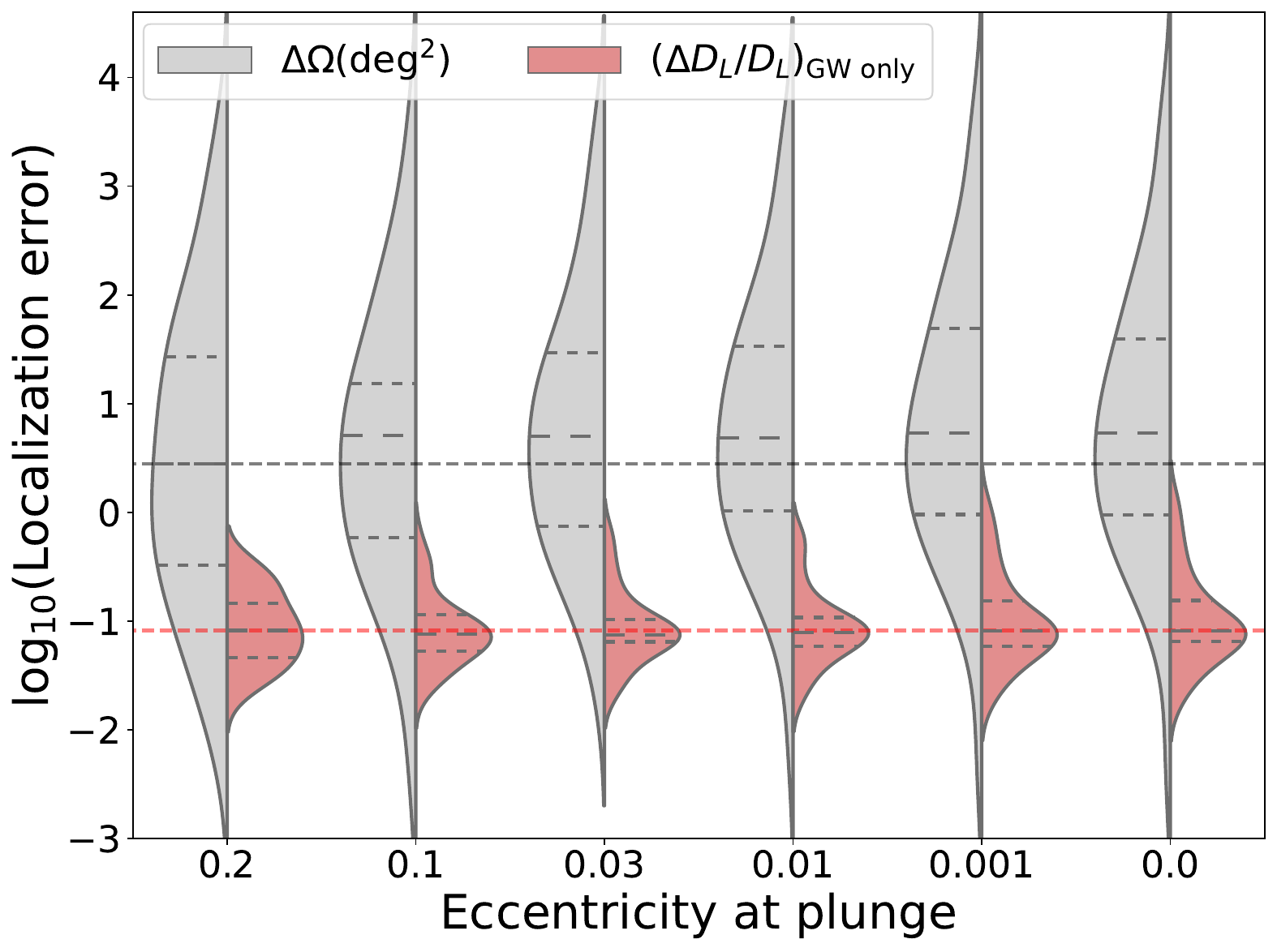}
\caption{Spatial localization errors 
of the TianQin EMRIs as a function of their eccentricities at the time of plunge. 
The gray (red) violinplots represent the probability distributions for 
the sky localizations errors ($D_L$ estimation errors) of EMRIs, 
and the gray (red) horizontal dashed line shows the median of the 
corresponding error. 
}
\label{fig:DLSkyerrs_eVary}
\end{figure}

In our simulations, we have set the eccentricities of EMRIs (at plunge) to be
uniformly distributed in the range of $0$ to $0.2$ when simulating the EMRIs
from normal galaxies, and set them to zero when the EMRIs come from AGNs 
\citep{2021PhRvD.103j3018P, 2022PhRvD.105h3005P, 
2021PhRvD.104f3007P, 2023LRR....26....2A}.
To evaluate whether the eccentricities of EMRIs may affect 
the precision in the measurement of their parameters,
here we analyze the spatial localization errors for the EMRIs
formed in different channels.
The results are shown in Figure~\ref{fig:DLSkyerrs_eVary}. 

From Figure \ref{fig:DLSkyerrs_eVary}, we can see that 
the effect of eccentricity at plunge is weak
on the
spatial localization errors (sky position and distance). 
The weak dependence implies that 
neither the errors on the cosmological parameters presented in 
Section \ref{sec:result_cosmo} nor
the required minimum number of EMRIs for establishing 
the EMRI-AGN correlation presented in Section \ref{sec:result_astrop} 
is sensitive to our assumption of the distribution of EMRI eccentricities. 

Although we have not considered in detail in this work, a special case
worth mentioning is an EMRI whose eccentricity is close to unity.
Such a system may occasionally form in a normal galaxy, and it would radiate 
multiple GW bursts before transitioning into the inspiraling phase \citep{2006AIPC..873..284R, 2007MNRAS.378..129H, 
2008ApJ...675..604Y, 2013MNRAS.435.3521B, 2019PhRvD..99l3025A, 2020MNRAS.498L..61H, 
2022PhRvD.106l4028F}. The SNR of this extreme-mass-ration burst (EMRB) may exceed 
the detection threshold, allowing identification of individual bursts \citep{2006AIPC..873..284R, 
2008ApJ...675..604Y, 2020MNRAS.498L..61H, 2022PhRvD.106l4028F, 2023arXiv230815354W}. 
Detecting these burst prior to an EMRI signal can enhance 
the spatial localization of the EMRI by partially breaking the degeneracy between parameters
\citep{2004PhRvD..69h2005B, 2012PhRvD..86j4027M, 2023PhRvD.107d3009X}. 
This prospect deserves future investigation.

\subsection{Prospect of clarifying the Hubble tension}    \label{sec:discuss_EMRIvsH0tension}

Recently, the cosmological standard model---flat-$\Lambda$CDM, is challenged by the
Hubble tension \citep{2017NatAs...1E.121F, 2021CQGra..38o3001D,
2022NewAR..9501659P}.  The $H_0$ obtained from the
flat-$\Lambda$CDM model plus cosmic microwave background is $67.4
\pm 0.5 ~\! {\rm km} ~\! {\rm s}^{-1} ~\!\! {\rm Mpc}^{-1}$, with a relative
error of about $0.7\%$ \citep{2020A&A...641A...6P}, and the $H_0$
from the cosmic distance ladders and type Ia supernovae 
is  $73.2 \pm 1.3 ~\! {\rm km} ~\! {\rm s}^{-1} ~\!\! {\rm
Mpc}^{-1}$, with a relative error of about $1.8\%$ \citep{2021ApJ...908L...6R}.
So if we want to clarify the Hubble tension by a new independent measurement,
the independently measured $H_0$ needs to be better than a precision of about
$2\%$. 

We can see from Table \ref{tab:constraint_errs} that TianQin does not achieve a
precision of 2\% in constraining $H_0$ under the EMRI population models considered in this paper. 
As for
TianQin I+II, it achieves a precision better than $2\%$ only under the
population models with optimistic EMRI rates, such as the M7 and M12 models.
However, the multi-detector network formed by LISA and TianQin, i.e.,
TianQin+LISA, can much better constrain $H_0$.  For example, under most of the
population models, except for the M8 and M11 models, the TianQin+LISA network
can constrains $H_0$ to a precision of near or better than $2\%$.  Therefore,
our results support a joint observation of GWs by TianQin and LISA. 

\section{Conclusion}    \label{sec:conclusion}

In this work, we have investigated the potential of using 
EMRIs as dark sirens to constrain the 
cosmological parameters as well as 
infer the formation channel of EMRIs. 
We adopted 11 population models 
presented in \cite{2017PhRvD..95j3012B} to generate 
catalogs of EMRIs which are detectable by different space GW detectors
or detector networks.. 

(i) To constrain the cosmological parameters, we derived the redshift
probability distributions of detected EMRIs by matching their spatial
localization error volumes with galaxy survey catalogs, and then evaluated the
posterior probability distributions of the cosmological parameters via a
Bayesian analytical framework. 
Our results show that precise constraints on the
Hubble-Lema\^itre constant $H_0$ and the dark energy EoS parameter $w_0$ can be
obtained from EMRIs.  Under the fiducial population model M1, the precision of
$H_0$ can reach about $8.1\%$ for TianQin, about $4.4\%$ for TianQin I+II, and
about $1.9\%$ for the TianQin+LISA network; and the precision of $w_0$ can
reach about $37\%$ for TianQin, about $12\%$ for TianQIn I+II, and about
$6.5\%$ for the TianQin+LISA network.  Additionally, under the most optimistic
EMRI population model, TianQin can constrain $H_0$ and $w_0$ to a precision
better than about $3\%$ and about $10\%$, and the TianQin+LISA network can
constrain $H_0$ and $w_0$ to a precision better than about $1\%$ and about
$5\%$, respectively.

(ii) To constrain the formation channel of EMRIs, 
we adopted an AGN luminosity function to self-consistently generate the numbers 
of candidate host AGNs for the detected EMRIs, and 
then used the Poisson distribution to statistically test 
the spatial correlation between EMRIs and AGNs. 
We mainly found that 
the minimum number $N_{{\rm GW},3\sigma}^{\rm threshold}$ of the EMRIs 
required to establish the EMRI-AGN correlation 
decreases with increasing $f_{\rm agn}$. 
If $f_{\rm agn} = 1$
and the cosmological parameters are fixed,  the values of $N_{{\rm GW},3\sigma}^{\rm threshold}$ 
for the three detector configurations, namely TianQin, 
TianQin I+II, and TianQin+LISA,
are about $1300$, $450$, and $30$ respectively. 
Under the condition that the cosmological parameters vary freely within the priors, 
the values of $N_{{\rm GW},3\sigma}^{\rm threshold}$ 
become significantly larger, by a factor of about three or larger. 
However, in our simulated cases where the correlation between EMRIs and AGNs can be
statistically confirmed,  
a combination of the EMRI and AGN catalogs can significantly improve the constraints 
on the cosmological parameters. 

The cosmological constraints reported in this work appear to be somewhat weaker than those of 
LISA alone reported in \cite{2021MNRAS.508.4512L} \cite[also see][]{2008PhRvD..77d3512M}. 
For example, under the M1, M5, and M6 models, LISA with a four year mission 
is expected to constrain $H_0$ and $w_0$ to higher precisions than TianQin or TianQin I+II. 
This is mainly because the most sensitive band of TianQin 
is slightly higher than that of LISA, 
which makes the EMRI detection rate of TianQin much lower than LISA \citep{2017PhRvD..95j3012B, 2020PhRvD.102f3016F}. 
However, we would like to point out that currently there is 
large uncertainty in the mass function of MBHs especially at the low mass end 
\citep{2007ApJ...667..131G, 2012MNRAS.423.2533B, 2019ApJ...883L..18G, 2024ApJ...962..152H}. 
The GW signals radiated by the EMRIs that are formed by lighter MBHs, 
will have higher frequencies and are more easily detected by TianQin. 

Our results also demonstrate the prospect of using space-borne GW detectors to
statistically infer the formation channel of EMRIs.  It is previously known
that the intrinsic parameters of an EMRI, such as the eccentricity at plunge
\citep{2021PhRvD.103j3018P, 2021PhRvD.104f3007P} and the spin of the MBH
\citep{2020PhRvD.102l4054B}, can help us infer the formation channel of the
EMRI, but an unbiased extraction of the intrinsic parameters relies heavily on
an accurate EMRI waveform model \citep{2017PhRvD..96d4005C,
2018LRR....21....4A, 2021PhRvL.127d1102T}. Now the effectiveness of the
statistical framework proposed in this work mainly relies on the spatial
localization accuracies of EMRIs and the completeness of AGN catalogs. Since
the localization accuracies are mainly determined by the SNRs of the EMRIs and
the completeness of the AGN catalogs is mainly determined by the depths of AGN
surveys, the demand of accurate waveform model subsides to a secondary role.
In this sense, our statistical framework could serve as a useful complementary  means
of inferring the formation channel of EMRIs. 

\section*{Acknowledgements}

The authors would like to thank Linhua Jiang, Yuming Fu, 
Zhenwei Lyu, Zhen Pan and Tieguang Zi for very helpful discussions, 
thank the anonymous referee for many helpful comments,
and thank the Theoretical Astrophysical Observatory 
(TAO; \url{https://tao.asvo.org.au/tao/}) 
for providing access to the download of the mock galaxy catalogs. 
This work is supported by the National Key Research and Development Program of China (Grant No. 2021YFC2203002) and
the National Natural Science Foundation of China (Grant No. 11991053).
L.-G. Zhu is funded by China Postdoctoral Science Foundation (Grant No. 2023M740113), 
H.-M. Fan is supported by the Hebei Natural Science Foundation (Grant No. A2023201041), 
Y.-M. Hu is supported by the Natural Science Foundation of China 
(Grants No. 12173104 and No. 12261131504), 
and Guangdong Major Project of Basic and Applied Basic Research (Grant No. 2019B030302001), 
and J.-d. Zhang is supported by the Guangdong Basic and Applied Basic Research Foundation (Grant No. 2023A1515030116). 
This work is also supported by Key Laboratory of TianQin Project 
(Sun Yat-sen University), Ministry of Education. 
The computations in this work was performed on the High
Performance Computing Platform of the Centre for Life Science, 
Peking University.

\software{\textsf{numpy} \citep{2011CSE....13b..22V}, 
\textsf{scipy} \citep{2020NatMe..17..261V}, 
\textsf{emcee} \citep{2013PASP..125..306F, 2019JOSS....4.1864F}, 
\textsf{matplotlib} \citep{2007CSE.....9...90H}, 
\textsf{corner} \citep{2016JOSS....1...24F}, 
\textsf{GetDist} \citep{2019arXiv191013970L}, 
and \textsf{seaborn} \citep{Waskom2021}. 
}

\appendix

\section{Distributions of the likelihood ratio in the null and alternative
hypotheses}    \label{appendix:lambda_bg}

\begin{figure}[h]
\centering
\includegraphics[width=0.480\textwidth]{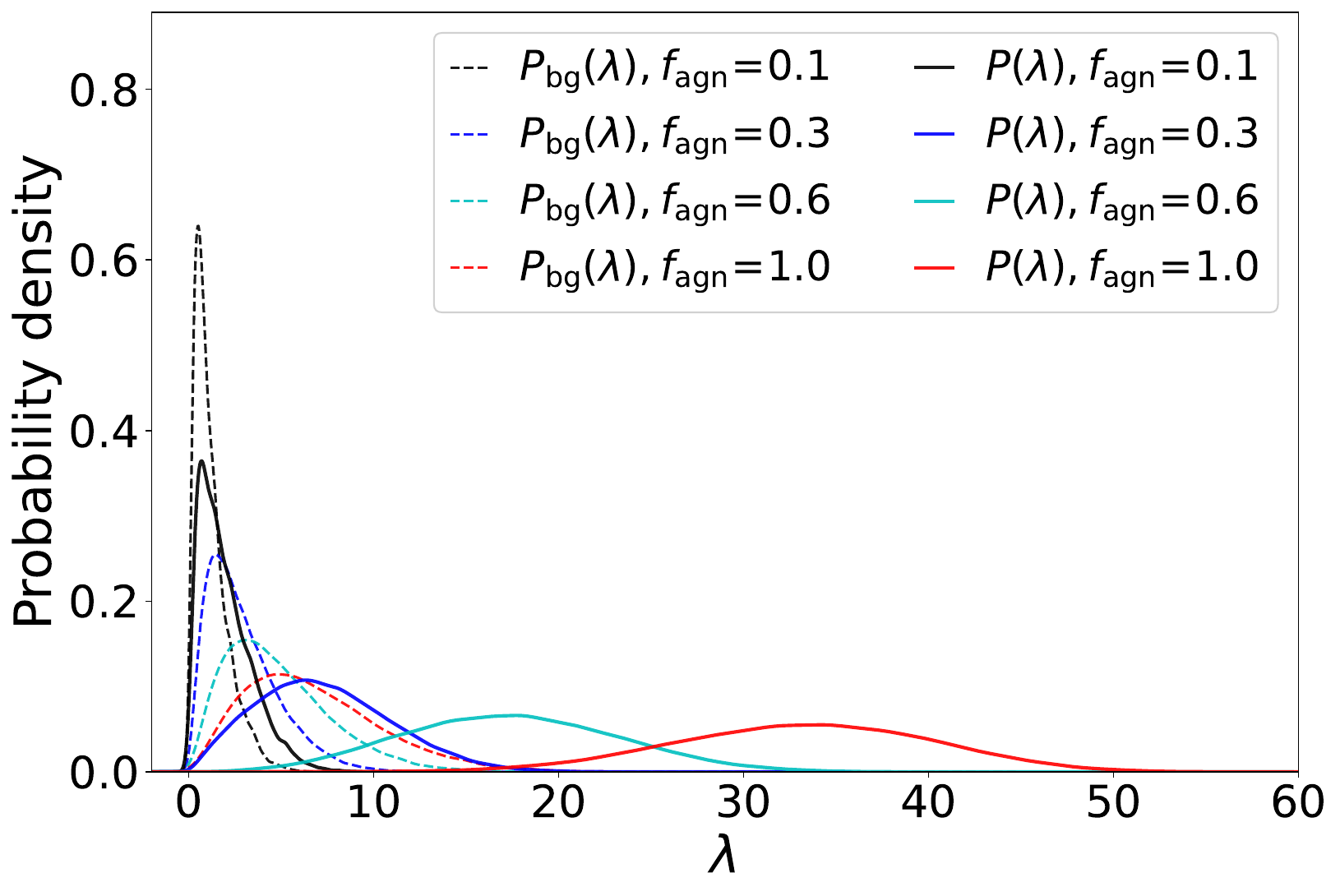}
\caption{Typical background distribution $P_{\rm bg}(\lambda)$ (dashed line) and 
probability distribution $P(\lambda)$ (solid line) of the likelihood ratio $\lambda$ 
in Equation (\ref{eq:likeli-ratio}) for 100 EMRIs detected by TianQin+LISA, 
assuming our fiducial cosmological parameters. 
The black, blue, cyan, and red colors correspond to four fractions of 
$f_{\rm agn} = 0.1, \!~0.3, \!~0.6, \!~1.0$, respectively. 
}
\label{fig:Pbg_P-lambda}
\end{figure}

In this appendix we illustrate a deviation of the distribution of the
likelihood ratio from the background distribution as $f_{\rm agn}$,  the
fraction of EMRIs originated in AGNs, increases.  We assume that $100$ EMRIs
are detected by the TianQin+LISA network, which is expected by most population
models for five years of observation, regardless of whether EMRIs originate
predominantly in normal galaxies or AGNs \citep{2017PhRvD..95j3012B,
2020PhRvD.102f3016F, 2021PhRvD.104f3007P}.  The background distributions
$P_{\rm bg}(\lambda)$ of the likelihood ratio are shown in Figure
\ref{fig:Pbg_P-lambda} as the dashed lines, while the probability distributions
$P(\lambda)$ of the ``signal'' likelihood ratio (derived from detected EMRIs)
are represented by solid lines 
(we wish to clarify that the signal likelihood ratio $\lambda$ is a deterministic function of $f_{\rm agn}$ 
in actual GW detections). 
Both distributions $P_{\rm bg}(\lambda)$ and
$P(\lambda)$ are derived from tens of thousands of independently repeated simulations.  We
can see that the variation of $P(\lambda)$ is more significant than $P_{\rm
bg}(\lambda)$ as $f_{\rm agn}$ increases. This discrepancy arises because the
variation of $P(\lambda)$ is caused by a combination of the increase in the
parameter $f_{\rm agn}$ and the resulting increasing number of candidate host
AGNs $\{ N_{{\rm agn},i} \}$, whereas the variation of $P_{\rm bg}(\lambda)$
solely stems from the change in the parameter $f_{\rm agn}$. 

\end{CJK*}

\bibliography{refs_emri}   

\end{document}